\documentclass[%
 aip,
 amsmath,amssymb,
 reprint,%
]{revtex4-2}
\usepackage[utf8]{inputenc}
\usepackage[T1]{fontenc}
\usepackage{mathptmx}
\usepackage{etoolbox}

\usepackage{bm}

\usepackage{graphicx}
\graphicspath{ {./figures/} {./} {./figures/ftr/} }

\usepackage{xcolor}

\usepackage{hyperref}
\hypersetup{
    colorlinks=true,
    linkcolor=blue,
    filecolor=magenta,      
    urlcolor=cyan,
}
\usepackage{floatrow}
\usepackage[caption=false]{subfig}
\usepackage{soul} 
\usepackage{ulem} 

\usepackage{fixme}
\fxsetup{draft}

\usepackage{todonotes}
\newcommand{\beq}{\begin{equation}}
\newcommand{\eeq}{\end{equation}}
\newcommand{\bea}{\begin{eqnarray}}
\newcommand{\eea}{\end{eqnarray}}
\newcommand{\p}{\partial}
\newcommand{\mb}{\mathbf}

\newcommand{\heinz}[1]{{\color{violet}#1}}
\newcommand{\andres}[1]{{\color{olive}#1}}

\makeatletter
\def\@email#1#2{%
 \endgroup
 \patchcmd{\titleblock@produce}
  {\frontmatter@RRAPformat}
  {\frontmatter@RRAPformat{\produce@RRAP{*#1\href{mailto:#2}{#2}}}\frontmatter@RRAPformat}
  {}{}
}%
\makeatother

\begin{document}

\title{Filamentary plasma eruptions and the heating and acceleration of electrons}

\author{Heinz Isliker} 
\email[Corresponding author: Heinz Isliker, email address: ]{isliker@astro.auth.gr}
\homepage{https://orcid.org/0000-0001-9782-2294}
\affiliation{Astronomy Lab, Dept. of Physics, Aristotle University of Thessaloniki, 541 24 Thessaloniki, Greece}

\author{Andres Cathey}
\homepage{https://orcid.org/0000-0001-7693-5556}

\author{Matthias Hoelzl}
\homepage{https://orcid.org/0000-0001-7921-9176}
\affiliation{Max Planck Institute for Plasma Physics, Boltzmannstr. 2, 85748 Garching b. M., Germany}

\author{Stanislas Pamela}
\homepage{https://orcid.org/0000-0001-8854-1749}\affiliation{CCFE, Culham Science Centre, OX14 3DB, UK}

\author{Loukas Vlahos}
\homepage{https://orcid.org/0000-0002-8700-4172}
\affiliation{Astronomy Lab, Dept. of Physics, Aristotle University of Thessaloniki, 541 24 Thessaloniki, Greece}

\begin{abstract}
    We present test-particle simulations of electrons during a nonlinear MHD simulation of a type-I edge localized mode (ELM) to explore the effect of an eruptive plasma filament on the kinetic level. The electrons are moderately heated and accelerated during the filamentary eruption on a fast time scale of the order of $0.5\,$ms. A clearly non-thermal tail is formed in the distribution of the kinetic energy that is of power-law shape and reaches 90 keV for some particles. The acceleration is exclusively observed in the direction  parallel to the magnetic field, i.e.\ with a clear preference in counter-current direction, and we show that the parallel electric field is the cause of the observed acceleration. Most particles that escape from the system leave at one distinct strike-line in the outer divertor leg at some time during their energization. The escaping high energy electrons in the tail of the energy distribution are not affected by collisions, they thus show characteristics of runaway electrons.
    The mean square displacement indicates that transport in energy space clearly is super-diffusive, and
    interpreting the acceleration process as a random walk, we find that the distributions of energy-increments exhibit exponential tails, and transport in energy space is
    equally important of convective (systematic) and diffusive (stochastic) nature. 
    By analyzing the MHD simulations per se, it turns out that the histograms of the parallel electric field in the edge region exhibit power-law shapes, and this clearly non-Gaussian statistics is ultimately one of the reasons for the moderately anomalous phenomena of particle transport that we find in energy space. 
\end{abstract}
\maketitle


\section{Introduction}

The formation of Coherent Structures (CoSs) (magnetic filaments, current sheets (CS), large amplitude magnetic disturbances, vortices and shocklets) in 2D and 3D strongly turbulent plasmas has been analysed extensively in the last forty years~\citep{Biskamp89, Dmitruk03, Arzner06, Servidio09, Servidio10,  Servidio11,  Zhdankin13, Valentini14, Karimabadi2014, Cerri17, Isliker17a, ComissoL19, Rueda21}.
CoSs appear intermittently inside turbulent plasmas and are collectively the locus of magnetic energy transfer (dissipation) into particle kinetic energy, leading to heating and/or acceleration.

In the early 80's, the link between reconnection and turbulence has been established~\citep{Matthaeus86}, and a few years later the link between turbulence and reconnection has also been analyzed~\citep{Biskamp89}. Several recent reviews discuss the way turbulence can become the host of reconnecting current sheets and how  reconnecting current sheets can drive turbulence~\citep{Matthaeus11, Cargill12, Lazarian12, Karimabadi2014, Vlahos19, Lazarian20}. 

There are several ways to initiate strong turbulence in 2D and 3D numerical simulations~\citep{Dmitruk03, Arzner06, Servidio11,  Zhdankin13, Valentini14, Karimabadi2014, Cerri17, Isliker17a, Rueda21}.   In these articles, the authors   did not set up a specific geometry of a reconnection environment or prescribe a collection of waves~\citep{Arzner04} as turbulence model, but allow the MHD equations themselves to build
naturally correlated field structures and
coherent regions of intense current densities (current filaments or CSs).
It is of foremost importance to find ways to identify 3D CoSs inside a turbulent plasma and measure their statistical characteristics. Several algorithms have been proposed in order to identify the geometrical structures of CoSs in numerical simulations and observations~\citep{Greco09, Servidio09, Servidio10,  Servidio11, Uritsky10,   Zhdankin13, Rossi15, DeGiorgio17, Fadanelli19,  Sisti21}. Notably,~\citet{Zhdankin13} were able to show that a large number of CSs in 3D turbulent environments do not contain reconnection sites, and likewise, many reconnection sites do not reside inside CSs. Large scale magnetic disturbances and CoSs in fully developed turbulence follow a monofractal or multi-fractal scaling, both in space and astrophysical plasmas~\citep{Tu95, Shivamoggi97, Biskamp00, Leonardis13, Schaffner15, Isliker19}. This information is very important for analysing the interaction of particles with CoSs~\citep{Sioulas20b}.

The fragmentation of a large scale CSs was analysed by several authors~\citep{Kowal11, Daughton11, Karimabadi13a, Oishi15, Dahlin15, Kowal17, Adhikari20}.  A different mechanism for a large scale CS to reach fragmentation may be the presence of other CoSs in the surrounding of the CS, e.g.\  multiple reconnection sites ~\citep{Burgess16}.

Turbulence and magnetic reconnection can be present in explosively unstable plasmas, forming a new electromagnetic environment, which can be called \textit{turbulent reconnection}, and where spontaneous formation of current sheets takes place.~\citet{Vlahos19} show that the heating and the acceleration of particles in such environments is the result of the synergy of stochastic (second order Fermi) and systematic (first order Fermi) acceleration inside fully developed turbulence.

Turbulence features predominantly also in tokamaks at different temporal and spatial scales across the plasma~\citep{Martines02, Galassi17, Galassi19}. In the plasma core, microscopic turbulence dominates and stiffly limits the steepness of the plasma density and temperature profiles~\cite{Garbet04}. In the plasma edge, i.e.\ near the magnetic separatrix (or last closed flux surface), microscopic and fluid turbulence can also be dominantly present during so-called low confinement regime (L-mode). However, strong suppression of turbulence leads to the formation of a transport barrier and to the onset of the high confinement regime (H-mode), which features a pedestal (across the edge transport barrier) that rigidly raises the core pressure~\citep{Wagner82,Burrell97}.
Edge localized modes (ELMs) are violent transient magnetohydrodynamic (MHD) instabilities that appear repetitively in standard tokamak H-mode plasmas due to large pressure gradients and current density in the pedestal region, located at the boundary of the plasma confined region. These edge localized instabilities lead to a periodic loss of the plasma confined in the edge region such that heat and particles get expelled on a time scale of ${\lesssim 1}$~ms, causing transient heat loads to divertor targets~\cite{Zohm1996,Connor1998,Leonard2014}. During the ELMs, magnetic field stochastization as well as filamentary eruptions are typically observed~\cite{Ham2020}.
Magnetic perturbations during ELMs are linked to reconnection as they result from resistive peeling-ballooning modes, which cause magnetic reconnection at the respective resonant surfaces. Overlapping islands then form a stochastic layer at the plasma boundary.
Fig.\ 13 of \citet{Huysmans2007} shows for the first time that an edge stochastic layer forms during ELMs, while
Fig.\ 1 of \citet{Dunne2012}
shows a "current spike" during the ELM crash, which is characteristic for reconnection events that are violent enough to change the current profile, and which can be seen as experimental evidence for a reconnection event. Similar current spikes are observed in disruptions.
Other references that support the connection of ELMs with reconnection include e.g.\  \citet{Cowley2003,Fundamenski2007,Ebrahimi2017,Galdon-Quiroga2018}.  

In present-day tokamaks, ELMs are not a cause for major concern, but for larger devices, like ITER, unmitigated ELMs are expected to cause a considerable reduction of the lifetime of divertor components~\cite{Loarte2003,Eich2017,Gunn2017}, 
and even mitigated ELMs might be intolerable for DEMO~\cite{Wenninger2015}. As a consequence, extensive research endeavors have been carried out in the past three decades to understand the dynamics of these edge localized instabilities. Of particular importance for the present work are the efforts from the theory and modeling research communities because in order to study the effect of ELMs on kinetic electron populations, it is necessary to have time-evolving magnetic and electric fields, which cannot be measured experimentally in the time-scales of interest with sufficient accuracy, but can be extracted from high-fidelity ELM simulations.

Numerical simulations of edge instabilities in tokamaks have improved significantly in the past decade~\cite{Huijsmans2015}, and have now reached a state in which even systematic validations with respect to experimental observations have been carried out~\cite{Pamela2017}. For the test-particles studies presented in this work, we make use of recent simulations of multiple repetitive type-I ELMs in the ASDEX Upgrade (AUG) tokamak~\cite{Cathey2020}. These type-I ELMs expel approximately $7\%$ of the plasma stored energy in time-scales on the order of one millisecond, and these losses were primarily caused by a stochastization of the plasma edge, which opens up an efficient loss channel along the magnetic field lines. More details regarding these simulations are presented in section~\ref{sect:MHD-model}.

Test-particle studies for the case of ions in MHD simulations of ELMs have been performed, e.g.\ in Refs.~\cite{Garcia-Munoz2013a,Garcia-Munoz2013b,Galdon-Quiroga2018}. In contrast, in the present paper we focus on electrons as test-particles, which on the kinetic level have been investigated in the MHD modelling of disruptions before to study formation and losses of runaway electrons \cite{Sommariva2018,Sommariva2018a,Tinguely2021}. 
In Ref.~\cite{Freethy2015}, measurements of microwave and x-ray emission during ELM activity at MAST were analyzed and combined with MHD and PIC simulations. It was found that there is acceleration of electrons dominantly in the direction parallel to the magnetic field, which though is followed by rapid isotropization, due to the action of the anomalous Doppler instability that causes fast collective radiative relaxation.
 

In this article, we perform test-particle simulations of electrons with the code MAGRA~\cite{Isliker2017a} in the time-evolving electromagnetic fields of an eruptive filament, simulated with the non-linear extended MHD code JOREK~\cite{Hoelzl2021,Huysmans2007,Czarny2008}, with focus on the dynamics in the edge region. The aim of the present work is to investigate the energy-dissipation of filaments at the kinetic level, addressing several issues: (1) We perform an analysis of the structure and statistics of the MHD data per se, to find hints about the cause of the kinetic phenomena on the larger-scale fluid level. (2)  We study the heating and acceleration of electrons during ELMs, and we also reveal the nature and true cause of the energization process. We also separately analyze the characteristics of the particles escaping from the system. (3) We investigate the nature of transport in energy-space, looking for non-Gaussian, anomalous transport phenomena, in several ways, and posing the question whether the classical Fokker-Planck paradigm applies, or whether a non-local, fractional transport approach is needed to describe the transport.  

The remainder of the article is structured as follows. In Sect.~\ref{sect:MHD-model}, we describe the model and setup of the MHD and the  test-particle simulations, respectively. Sect.~\ref{sect:MHD-data} contains the statistical analysis of the MHD simulations and also presents the spatial structure of the perturbations. In Sect.~\ref{sec:test-particles}, the results on the energization of the test-particle simulations are described, and the nature and cause of acceleration and heating are examined. Thereafter, 
in Sect.~\ref{sec:transport_energy}, the transport of the test-particles in energy space is analyzed, 
including an analysis of the nature of transport. Sect.~\ref{sect:collisions} scrutinizes the effects of collisions, and the conclusions are given in Sect.~\ref{sec:conclusions}.

\section{The model and the set-up}\label{sect:MHD-model}

\subsection{MHD simulations}\label{sec:MHD-simulations}

\begin{figure*}[htp]
	\centering
	\includegraphics[width=0.45\linewidth]
    {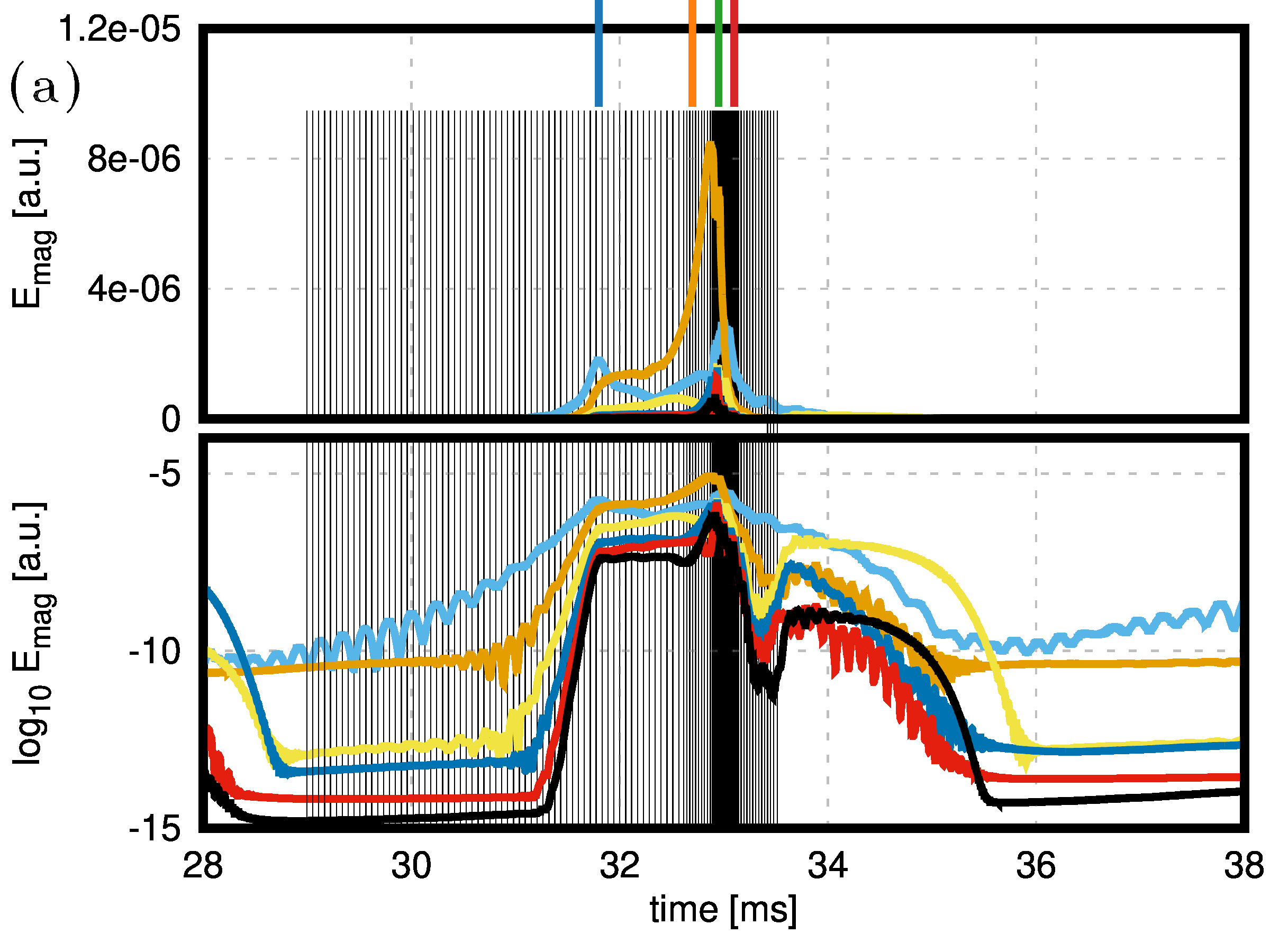}
	\includegraphics[width=0.45\linewidth]
    {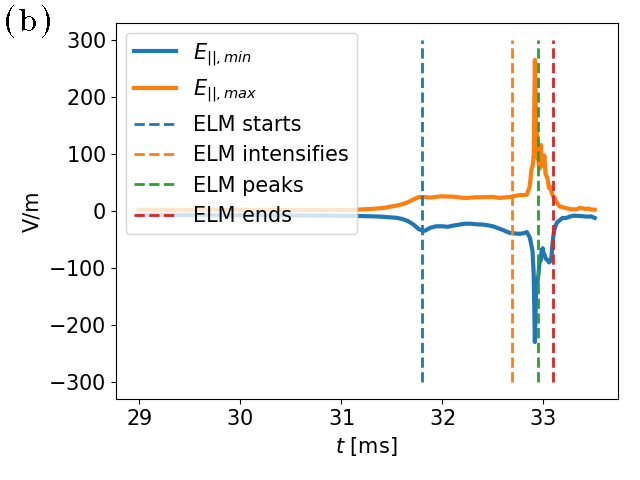}
	\caption{(a) Time evolution of the magnetic energies of the non-axisymmetric perturbations in linear (top) and logarithmic (bottom) scales. The black solid vertical lines indicate the times at which the JOREK data is output to then use in the test-particle simulations. (b) The minimum and the maximum of the parallel electric field inside of the confined plasma (i.e.\ within the separatrix) as a function of time. In both figures, the colored vertical lines mark the times of interest listed in Table \ref{tab:toi}. }
	\label{fig:ELM_evolution}
\end{figure*}

\begin{table}
    \centering
	\begin{tabular}{|c|l|l|}
		\hline
		$t_1$ & 31.80 ms  & ELM starts \\
		\hline
		$t_2$ & 32.70 ms  & ELM intensifies \\
		\hline
		$t_3$ & 32.95 ms & ELM peaks \\
		\hline
		$t_4$ & 33.10 ms  & ELM ends \\
		\hline
	\end{tabular}
	\caption{Times of interest for the ELM evolution.  }
	\label{tab:toi}
\end{table}

The test-particle studies performed in this article are based on simulations of type-I ELM crashes performed with the non-linear extended MHD code JOREK. Generally, the code evolves reduced or full MHD equations in a fully implicit time-stepping scheme for realistic tokamak X-point geometries using a 2D finite element grid combined with a toroidal Fourier representation. The ELM crash considered for the studies presented here corresponds to the third ELM from a simulation of four consecutive type-I ELMs~\cite{Cathey2020}. This simulation was performed with the single temperature reduced MHD model of JOREK, which is described in detail in Refs.~\cite{Franck2015,Hoelzl2021}. The simulation was initialised based on an equilibrium reconstruction from an ASDEX Upgrade (AUG) discharge with low triangularity, ${B_t=2.5~\mathrm{T}}$, ${I_p=0.8\mathrm{MA}}$, ${q_{95}\approx5.8}$, and high separatrix density ${(n_\mathrm{sep}\sim 0.4~\mathrm{n_\mathrm{GW}})}$, where ${n_\mathrm{GW}}$ is the Greenwald density. The toroidal mode numbers included in the simulation were ${n=0,2,4,\dots,12}$; it was not possible to include the odd mode numbers as well because an ${m/n=2/1}$ tearing mode was unstable and would interfere with the ELM dynamics.

One important simplification in the reduced MHD model is that the toroidal magnetic field is stationary and, as such, the time-varying poloidal magnetic field can be expressed with the toroidal component of the magnetic vector potential, ${\bm B_\mathrm{pol} = \bm\nabla \times \bm A_\varphi}$ 
. Faraday's law can then be used to express the electric field as, ${\bm E = - \partial_t A_\varphi \hat{\bm\varphi} - \bm\nabla \Phi}$, where ${\Phi}$ is the electrostatic potential. The JOREK electric field that is given as input for the MAGRA simulations is thus mainly determined by the perturbations to the magnetic vector potential caused by the ELM precursors and, in an even stronger manner, by the ELM crash. These perturbations are associated with reconnection and ergodisation of the magnetic fields in the plasma edge.  

The evolution of the self-generated bootstrap current is obtained in the simulation by making use of the analytical Sauter formulae~\cite{Sauter1999,Sauter2002}, and the evolution of the radial electric field $E_r$ (particularly, the $E_r$ well that is characteristic of H-mode plasmas~\cite{Viezzer2013}) is recovered by including the two-fluid effect of the diamagnetic drifts~\cite{Orain2015}. The latter two considerations are important to retain, since the current density is an important destabilising agent for the low-$n$ peeling modes (where $n$ is the toroidal mode number)~\cite{snyder2002edge}, and because the diamagnetic drift and the radial electric field play an important role in stabilising high-$n$ ballooning modes~\cite{rogers1999diamagnetic,hastie2000effect,huysmans2001modeling}. In the pedestal region, the resistivity considered for the simulation lies within the experimental error bars of the neoclassically-corrected Spitzer-resistivity. For further details on this simulation the reader is referred to Ref.~\cite{Cathey2020}.

In the type-I ELM simulation, four consecutive ELM crashes were modeled, and for the present work we consider the third of these ELMs, which starts to takes place at ${t=31.80~\mathrm{ms}}$.  As mentioned, this simulation was carried out with all even toroidal mode numbers between $n=0$ and $n=12$, and the magnetic energies of the non-axisymmetric perturbations corresponding to these toroidal mode numbers are plotted in linear and logarithmic scale in Fig.~\ref{fig:ELM_evolution}(a). In addition to the magnetic energies, vertical lines corresponding to the times for which JOREK data is extracted are also included, being more frequent around the ELM's peak for better dynamic resolution. In addition, Fig.~\ref{fig:ELM_evolution}(b) shows the minimum and maximum of the parallel electric field inside of the confined plasma (namely, inside of the separatrix). Table \ref{tab:toi} summarizes the four basic times of interest that we will consider throughout the following: the onset of the ELM activity at $t_1$, the start of the violent ELM crash at $t_2$, its peak at time $t_3$, and its end at $t_4$. The test-particles will be injected just before the onset of the violent ELM crash (at time $t_2$) and will be followed slightly beyond time $t_4$.

\subsection{Test-particle simulations\label{sec:TP_setup}}

\begin{figure}[htp]
		\centering
		\includegraphics[width=0.9\linewidth]{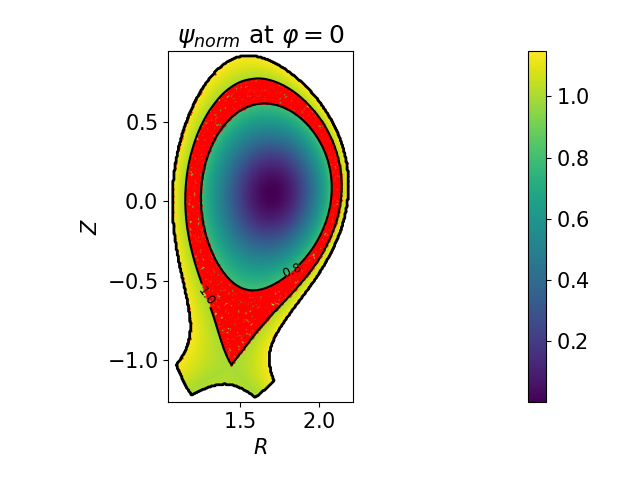}
		\caption{Pseudocolor plot of the normalized flux $\psi_{norm}$ from the MHD simulations in the $\varphi=0$ plane and at the initial time $t=32.5\,$ms of the test-particle simulations, with a few contour-lines, including $\psi_{norm}=1$ (separatrix) and $\psi_{norm}=0.8$.
		The particles are initialized within the red area with a random $\varphi$ location. }
	\label{fig:ini_cond}
	\end{figure}

\begin{figure*}[htp]
	\centering	
	\includegraphics[width=0.45\linewidth]{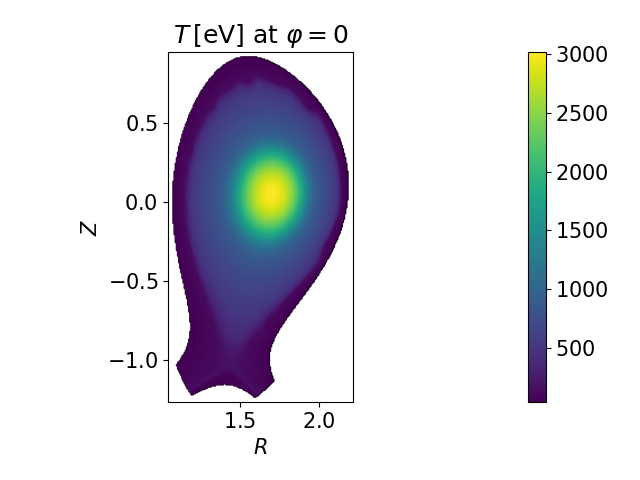}
	\includegraphics[width=0.45\linewidth]{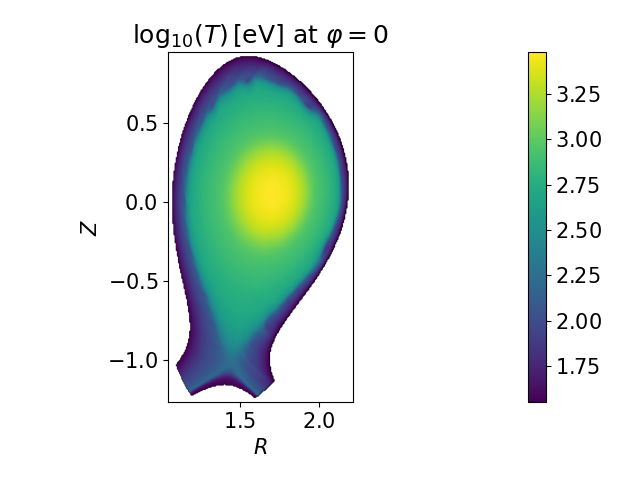}
	\caption{Left: Electron temperature $T$ from the MHD simulation with JOREK in the $\varphi=0$ plane, at the initial time $t=32.5\,$ms of the test-particle simulations. Right: The same data is plotted logarithmically to make perturbations close to the plasma boundary visible.
	}
	\label{fig:2}
	\end{figure*}

    The test-particle code MAGRA~\cite{Isliker2017a} solves the relativistic guiding center (GC) equations (here without collisions, see Sect.\ \ref{sect:collisions}) 
	for the evolution of the position $\mb{r}$ and the parallel component $u_{||}$ of the relativistic 4-velocity ${\bf u}=\gamma{\bf v}$ of the particles~\cite{Tao2007} 
	\beq
	\frac{d\mb{r}}{dt}= \frac{1}{B_{||}^*} 
	\left[ \frac{u_{||}}{\gamma} \mb{B}^* + \hat{\mb{b}}\times 
	\left(\frac{\mu}{q\gamma}\nabla B -\mb{E} \right) \right] ,
	\label{dr_dt}
	\eeq
	\beq
	\frac{du_{||}}{dt} = - \frac{q}{m_0 B_{||}^*}\mb{B}^* 
	\cdot\left(\frac{\mu}{q\gamma} \nabla B -\mb{E} \right) ,
	\label{dupar_dt}
	\eeq
	where 
	$\mb{B}^*=\mb{B} +\frac{m_0}{q} u_{||}\nabla\times\hat{\mb{b}}$,
	$\mb{B}$ is the magnetic and $\mb{E}$ the electric field,
	$\mu   = \frac{m_0 u_\perp^2}{2 B}$ 
	is the magnetic moment, 
	$\gamma=\sqrt{1+\frac{u^2}{c^2}}$,
	$B=|\mb{B}|$, $\hat{\mb{b}}=\mb{B}/B$, 
	$u_\perp$ is the perpendicular component of the relativistic 4-velocity,
	and $q$, $m_0$ are the particle charge and rest-mass, respectively. 
	Note that the term of the effective electric field proportional to $\frac{\partial{\bf b}}{\partial t}$ in \citet{Tao2007} is already included in the expression for the electric field $\mathbf{E}$ that is output from JOREK, see Sect.\ \ref{sec:MHD-simulations}.
	The GC equations are numerically integrated with an adaptive step-size Runge-Kutta/Dormand-Prince scheme.
	Note that similar options for test-particle simulations exist directly in JOREK \cite{Sarkimaki2022a}, and calculations by both codes have been compared to ensure correctness of the data transfer.
	
	In MAGRA, 3D local cubic interpolation (continuous in the values and some derivatives, i.e.\ tricubic or Hermite interpolation) of the values of the electromagnetic fields \textbf{B} and \textbf{E} on the spatial grid used by MAGRA (see below) is applied to determine the fields at the actual particle position in space.
	The parallel component $E_{||}$
	of the electric field is calculated by JOREK and interpolated separately by MAGRA, in order to avoid numerical errors when compared to the  values on the grid, which is a necessary procedure that has well been tested in several other applications \cite{Isliker17a,Isliker19}.

	The test-particles and the MHD perturbations evolve on similar time-scales, so that we use a time-series of MHD frames in the test-particle simulations. From the MHD simulations, there are 43 frames of JOREK-data for the modeling of the peak activity of the filament eruption, with a total duration of $0.7\,$ms, which naturally coincides with the total duration  of the test-particle simulations. Linear interpolation is applied in time direction between subsequent frames of JOREK data.
		
	For the test-particle simulations, the electromagnetic fields are interpolated from JOREK's internal flux surface-aligned grid to uniformly spaced cylindrical coordinates $(R,\varphi,Z)$, with spatial resolution of $0.7\,$cm in the $R$ and $Z$ direction, and angular resolution $2\pi/64$ in the $\varphi$ direction. MAGRA's cylindrical grid contains JOREK's  entire grid in the $R$-$Z$-plane (the colored part in Fig.\ \ref{fig:ini_cond}), it is though larger since it is of rectangular shape (the entire rectangle of Fig.\ \ref{fig:ini_cond}), and for practical reasons cylindrical grid-points outside the domain of JOREK's grid are assigned unphysical values such that MAGRA can detect when the simulation domain is left by a particle trajectory. 
		
	The normalized poloidal flux \(\psi_{norm}\) (0 at the magnetic axis and 1 at the separatrix) is used to determine the initial and the stopping conditions. The leaving particles are stopped as close as possible to the edge of the simulation volume, when they  enter an elementary cylindrical grid cube of which at least one node lies outside the modeling region, as signaled by an unphysical $\psi_{norm}$-value at a node.
				
	Spatial initial conditions are chosen based on \(\psi_{norm}\) and lie just inside the separatrix, in the entire \(R\)-$Z$-range corresponding to
	\(0.8\leq \psi_{norm} < 1\), uniformly distributed over the intended poloidal area by means of a rejection method, with  \(\varphi\) uniformly random in \([0,2\pi]\), as illustrated by Fig.\ \ref{fig:ini_cond}.

	The initial velocity distribution of the test-particles is assumed to be a Maxwellian (i.e.\ the three velocity components independently follow  Gaussian distributions), and the temperature of the Maxwellian distribution is determined through 3D linear interpolation of JOREK's electron-temperature field \(T_e\) (see Fig.\ \ref{fig:2}) to the initial particle position. The mean initial temperature in this region equals $0.46\,$keV. The direction of the initial velocity is chosen as uniformly random. For the statistical results that we present on from Sec.\ \ref{sec:test-particles}, we throughout consider a total of 100'000 test-particles.

\section{Spatio-temporal structures and statistical properties of the filaments \label{sect:MHD-data}}

\begin{figure*}[htp]
    \centering
    \includegraphics[width=0.45\linewidth]{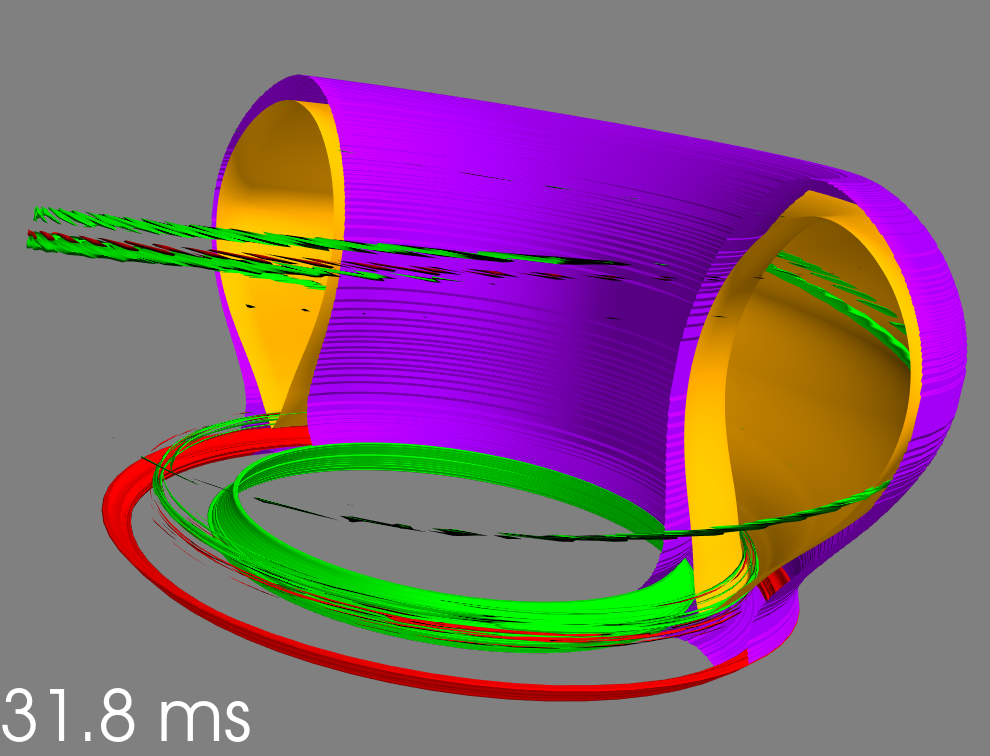}
    \includegraphics[width=0.45\linewidth]{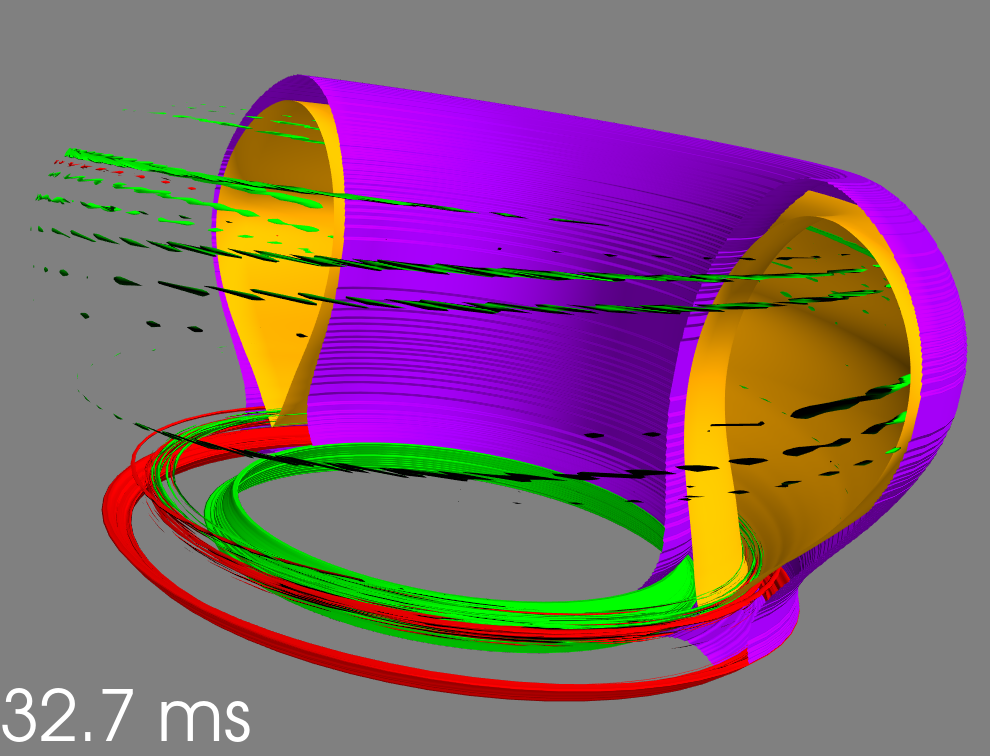}
    \includegraphics[width=0.45\linewidth]{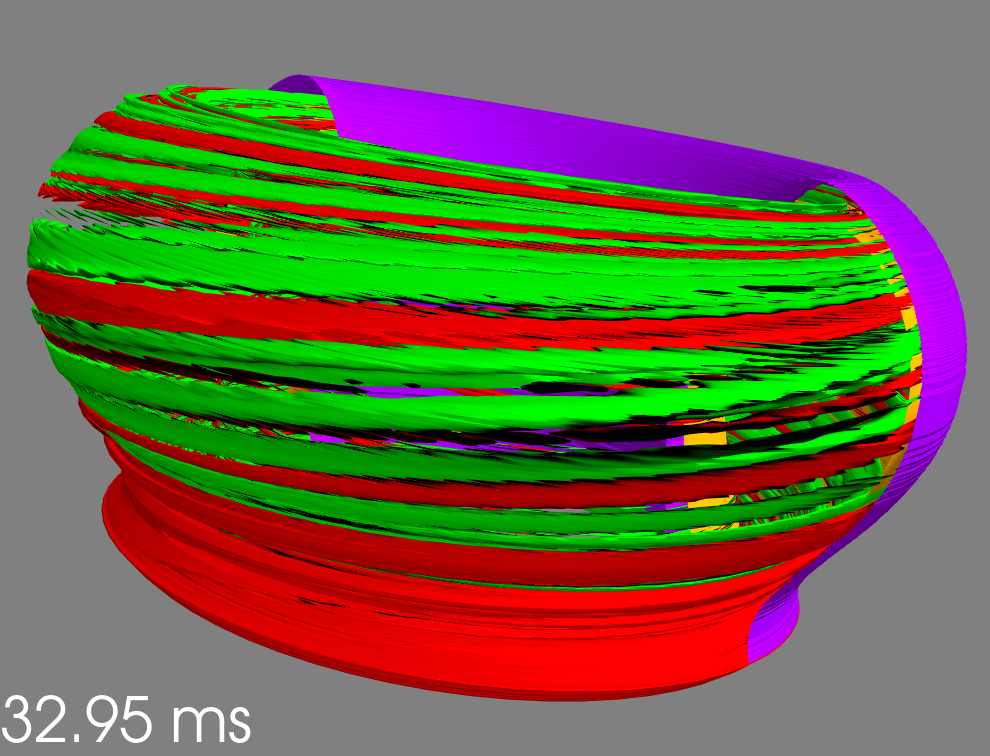}
    \includegraphics[width=0.45\linewidth]{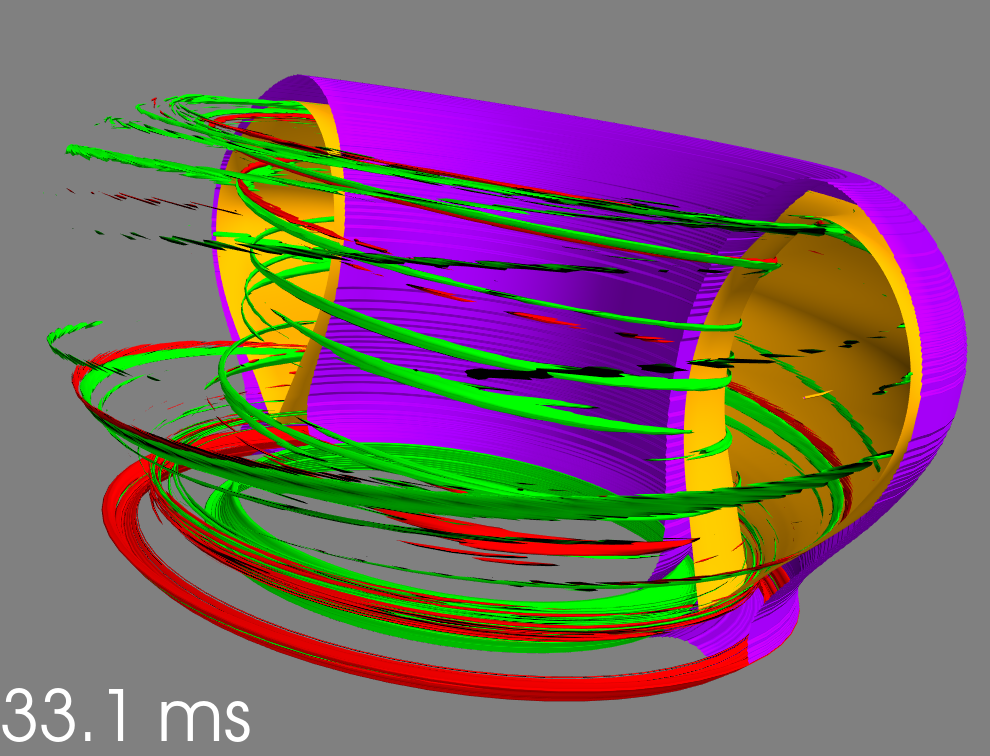}
    \caption{Iso-contours of $E_{||}$ (positive at +10 V/m in red, and negative at -10 V/m in green), for the times of interest $t=t_1,\,t_2,\,t_3,\,t_4$ (see Table \ref{tab:toi}), together with the separatrix (orange) and the plasma-boundary (violet) (both surfaces are half cut out).
    }
    \label{fig:Epar_3D}
\end{figure*}

\begin{figure*}[htp]
    \centering
    \includegraphics[width=0.45\linewidth]{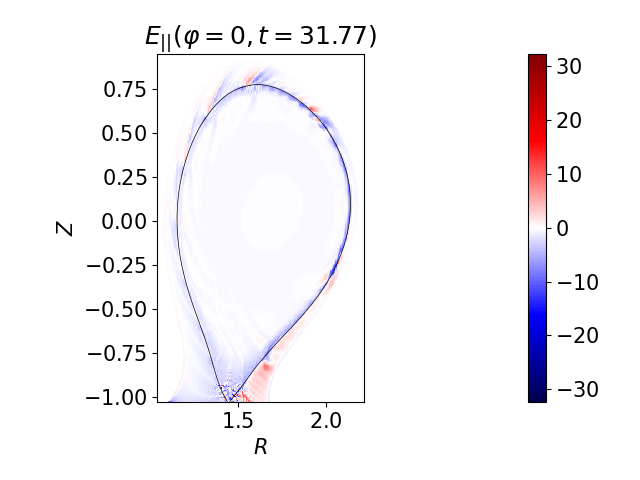}
    \includegraphics[width=0.45\linewidth]{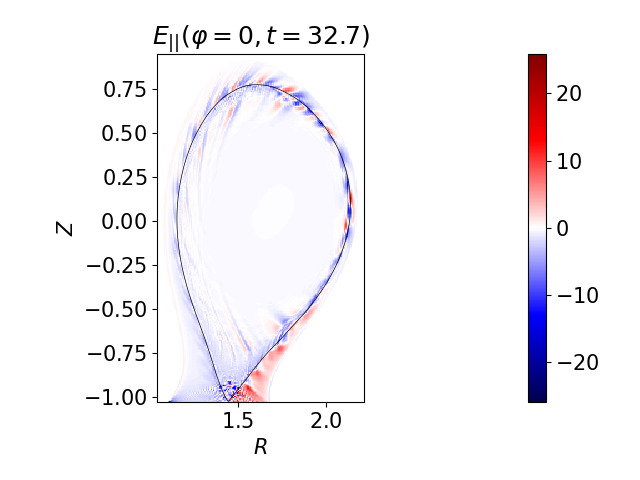}
    \includegraphics[width=0.45\linewidth]{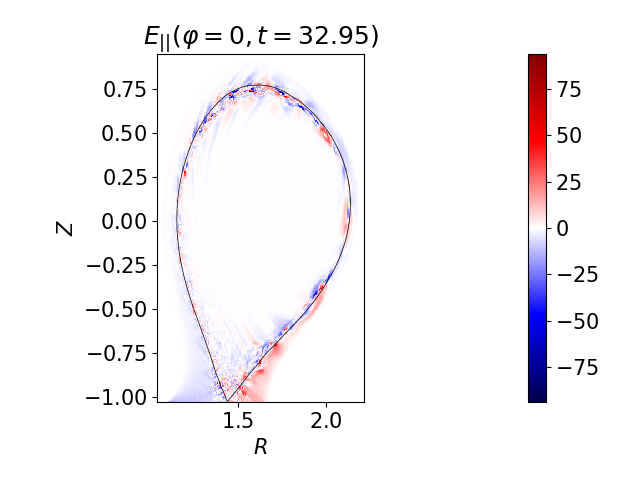}
    \includegraphics[width=0.45\linewidth]{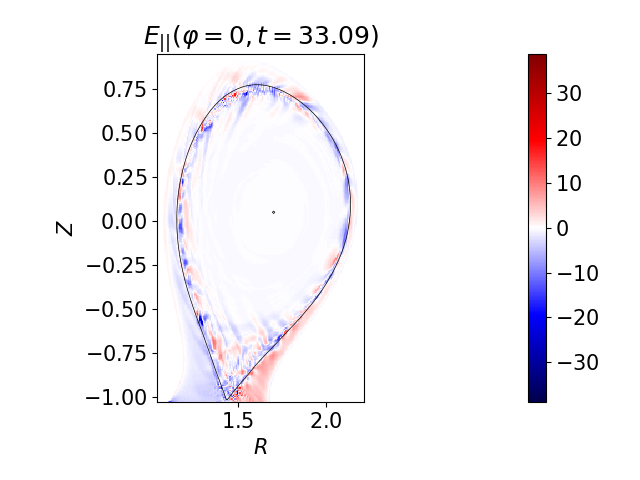}
    \caption{Pseudo-color representation of $E_{||}$ in the $\varphi =0$ plane, for the times of interest $t=t_1,\,t_2,\,t_3,\,t_4$ (see Table \ref{tab:toi}), together with the separatrix (black). For the sake of better visualization, the entire divertor region below the X-point is omitted, and a symmetric color range has been chosen such that $E_{||}=0$ is in white color. }
    \label{fig:Epar_2D}
\end{figure*}

In this section, which presents  the spatio-temporal properties of filaments, we focus on the parallel electric field, since it  is the ultimate cause for the particle acceleration  that we report and analyze in Sec.\ \ref{sec:test-particles} below. 

Fig.\ \ref{fig:Epar_3D} shows iso-contours of the parallel electric field $E_{||}$, at a positive and negative level, chosen for the sake of a clear visualization.
Band-like helical structures appear at 
the low-field side (LFS),  
as well as at the high-field side (HFS) when close to the peak activity of the ELM, where the structures also are larger in number and more dense, and they reach higher field values. The filamentary structures throughout appear in close vicinity of the separatrix (as it is also visible in the 2D  poloidal representations of the parallel electric field in Fig.\ \ref{fig:Epar_2D}, which reveal richly detailed structures).
Enhanced parallel electric fields are also present at the legs of the divertor region,
where the parallel electric field in any case attains its highest values (of the order of $\pm 1000\,$V/m at the bottom of the divertor legs during the ELM's peak). 

\begin{figure}[htp]
    \centering
    \includegraphics[width=0.9\linewidth]{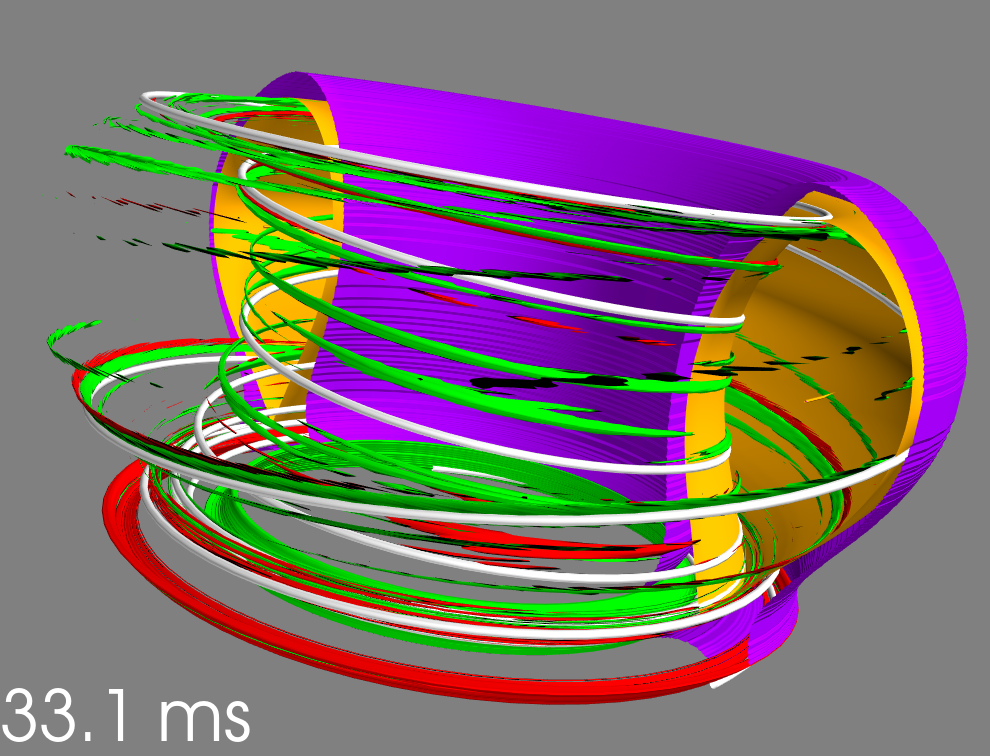}
    \caption{Same as Fig.\  \ref{fig:Epar_3D} for $t=t_4$, with additionally a field-line traced for a few toroidal turns shown in white color.  }
    \label{fig:Epar_3D_fl}
\end{figure}

Fig.\ \ref{fig:Epar_3D_fl} shows again iso-contours of the parallel electric field, as in Fig.\ \ref{fig:Epar_3D}, for one time-instance and together with a magnetic field line randomly chosen near the plasma-edge at the LFS and traced for several toroidal turns.
The figure illustrates that the helical structures are aligned with the magnetic field, as expected for MHD activity. 

We may conclude that the structures of
the parallel electric field trace the eruptive filamentary activity (the parallel electric field is actually correlated with density fluctuations, see e.g.\ Fig.\ 4 in~\cite{Freethy2015}).


\begin{figure}[htp]
	\centering
	\includegraphics[width=0.9\linewidth]{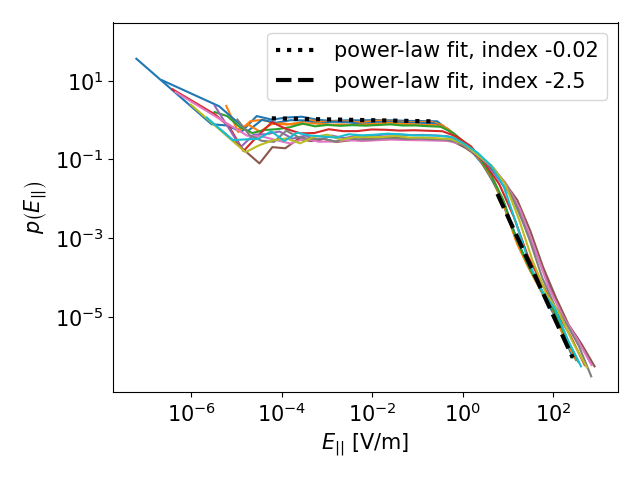}
	\caption{Histograms $p\left(| E_{||} | \right)$ of the magnitude of the parallel electric field $| E_{||} |$ in regions where $\psi > 0.8$ (edge region), for each MHD frame separately, in the range $t_1 < t < t_4\,$ms (see Table \ref{tab:toi}), together with two power-law fits to a distribution at one time-instance. }
	\label{fig:histoepar}
\end{figure}

\begin{figure*}[hbtp]
	\centering
	\includegraphics[width=0.45\linewidth]{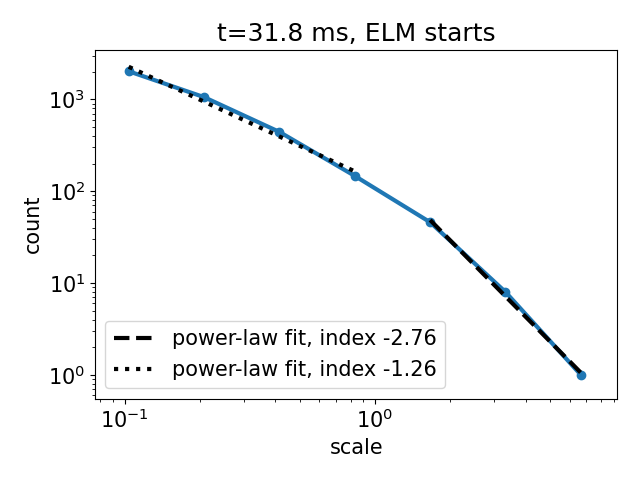}
	\includegraphics[width=0.45\linewidth]{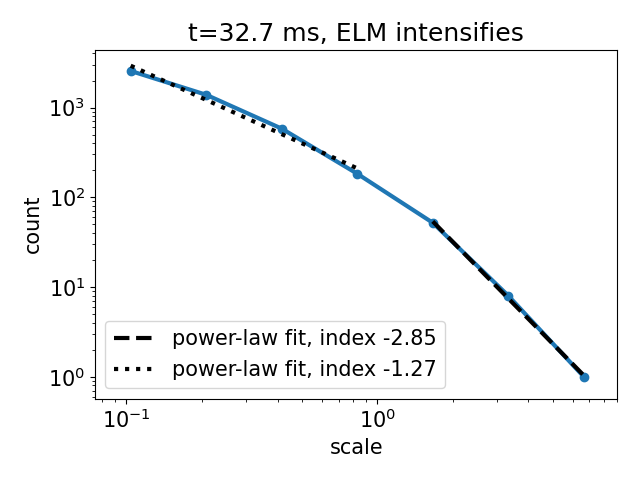}
	\includegraphics[width=0.45\linewidth]{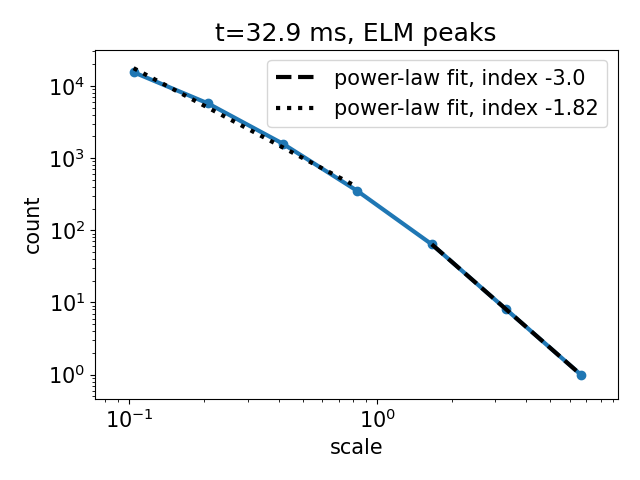}
	\includegraphics[width=0.45\linewidth]{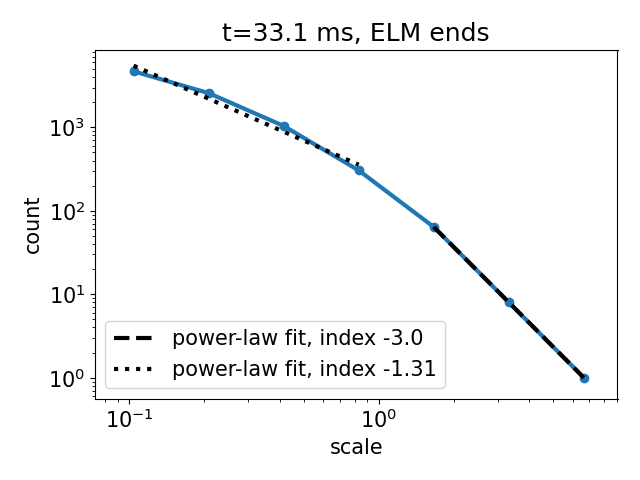}
	\caption{Fractal dimension estimate with the box-counting method: box-count as a function of box-size, for the four times of interest (see Table \ref{tab:toi}), together with power-law fits.
	}
	\label{fig:fractaldim}
\end{figure*}

Fig.\ \ref{fig:histoepar} shows the histogram $p\left(| E_{||} | \right)$ of the magnitude of the parallel electric field $| E_{||} |$ in the edge region (sampled from all grid-points located at $\psi > 0.8$), calculated and shown separately for each time-instant of JOREK output data in the time-range $t_1 < t < t_4\,$.
The histograms are similar in shape for all frames shown (and rather insensitive to the applied spatial threshold $\psi > 0.8$). They are close to flat at low values, and quite close to a power-law shape at high values, with power-law index $-2.5$.
The histograms thus   clearly obey non-Gaussian statistics, which is one of the ultimate reasons for the phenomena of particle acceleration that we find and present in Sec.\ \ref{sec:test-particles}.

We now turn to the question whether the structures in the parallel electric field, as shown in Fig.\ \ref{fig:Epar_3D}, are of fractal nature, applying 
the 3D box-counting method (e.g.~\cite{Falconer1990}) in order to determine the fractal dimension.
We consider regions of $|E_{||}| > 10$ V/m (the contour-level applied in Fig.\  \ref{fig:Epar_3D}), and the results from the box-counting algorithm are shown in Fig.\ \ref{fig:fractaldim}, for the four different times of interest defined in Table \ref{tab:toi}. There are in any case two distinct scaling regions of clear power-law shape. At small scales, a power-law fit reveals a fractal dimension close to one, except for the time at which the ELM peaks, where it is close to two.  At large scales, the fractal dimension is rather close to three in all cases. This implies that on a close view (at small scales) the structures are close to line- or tube-like and they get sheet-like when the ELM peaks, and on a global view (at large scales) they are 
almost space-filling, all in accordance with the visual impression given by Fig.\ \ref{fig:Epar_3D}. 
We thus may say
that the filaments cannot be considered to form a fractal structure. 

The applied threshold of $|E_{||}|>10\,$V/m was chosen for the sake of a clear visualization in Fig.\ \ref{fig:Epar_3D}. Lowering the threshold leads to space filling at increasingly smaller scales, while raising the threshold causes a thinning of the tube-  or sheet-like filaments until they disappear completely for too high thresholds.
In other words, changing the threshold within an appropriate range does not lead to noteworthy changes of the estimated values of the fractal dimension, it only shrinks or enlarges the extent of the power-law scaling ranges in Fig.\ \ref{fig:fractaldim}. 

\section{Test-particle simulations \label{sec:test-particles}}

\begin{figure}[htp]
    \centering
    \includegraphics[width=0.9\linewidth]{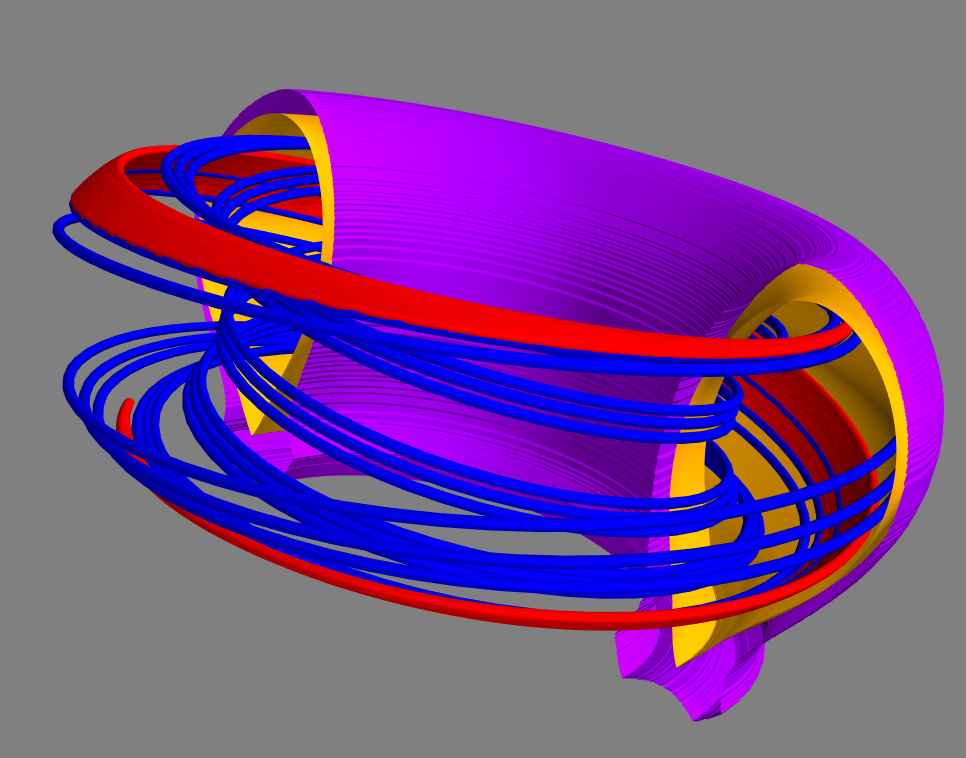}
    \caption{Separatrix (yellow), domain boundary (violet), and part of two typical particle orbits in space, a passing one (blue, the first 2\% of the integration time are shown), and a trapped one (red, the first 20\% of the integration time are shown).  }
    \label{fig:orbits}
\end{figure}

In the following, we throughout consider electrons, and we show results from tracking 100'000 test-particles from the time $t_0=32.5\,$ms up to the final time $t_f = 33.2\,$ms (duration $0.7\,$ms), i.e.\ from slightly before the time of interest $t_2$ (the ELM intensifies) until just after the time of interest $t_4$ (the ELM ends), see Table \ref{tab:toi} and Fig.\ \ref{fig:ELM_evolution}. We thus kinetically explore the most intense ELM phase.
As an illustration for the simulations, Fig.\ \ref{fig:orbits} shows part of two typical particle orbits in space,
a passing one (blue) that gets very close to the separatrix, and a trapped one (red). 

\subsection{Energization}
\label{sec:energization}

\begin{figure*}[htp]
	\centering
	\includegraphics[width=0.45\linewidth]{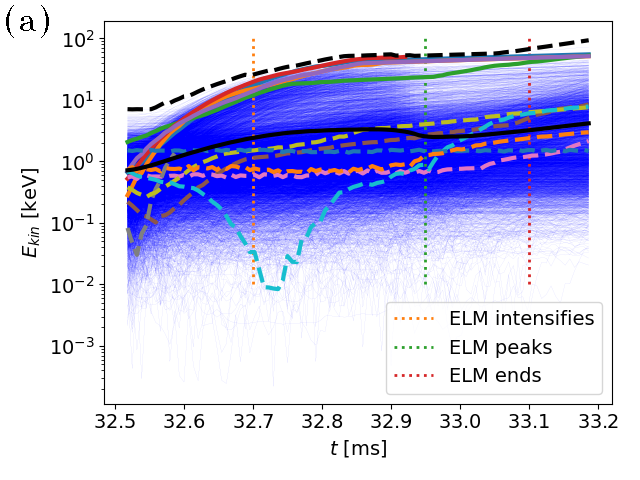}
	\includegraphics[width=0.45\linewidth]{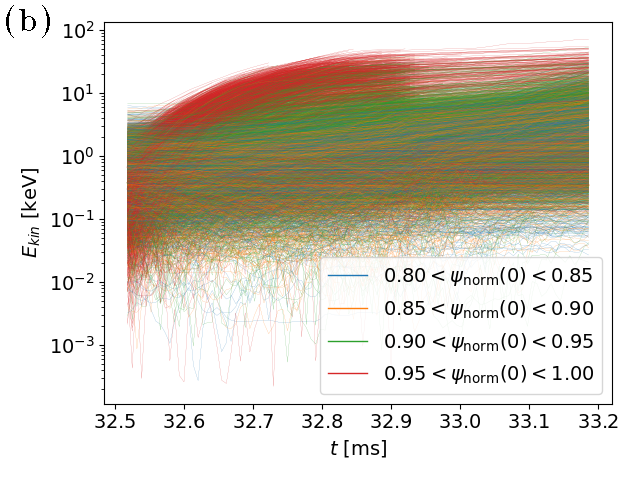}
	\\
	\includegraphics[width=0.45\linewidth]{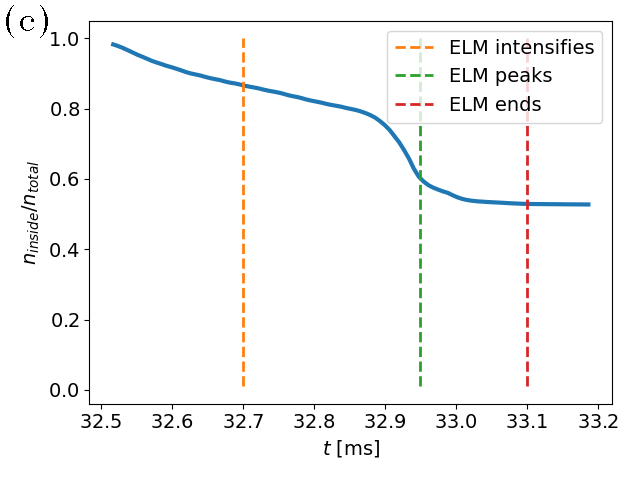}
	\\
	\includegraphics[width=0.45\linewidth]{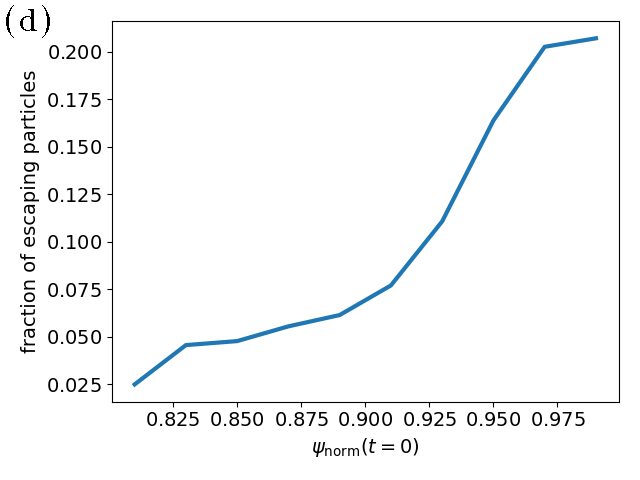}
	\includegraphics[width=0.45\linewidth]{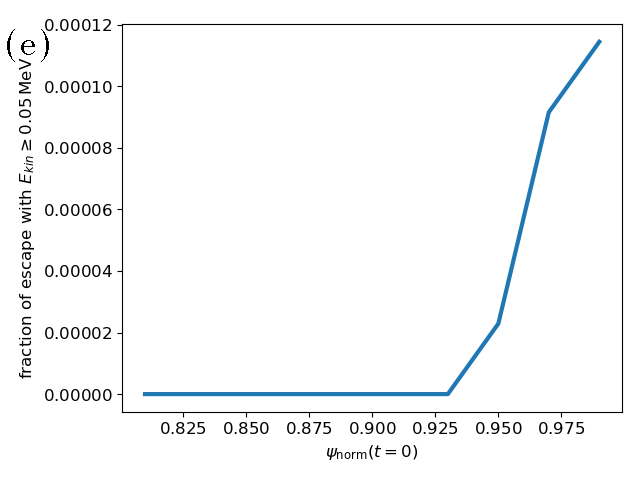}
	\caption{(a) Particle kinetic energies (thin blue), mean kinetic energy (thick solid black), and maximum kinetic energy (thick dashed black) of the particles not yet lost from the system as a function of time; the orbits of a few energetic particles (solid), and of a few low energy particles (dashed) have been marked by thick colored lines. --- (b) Particle kinetic energies colored according to the initial value of the normalized flux $\psi_\mathrm{norm}$. 
	--- (c) Number of particles (normalized to one) that still are in the system as a function of time.
	--- (d) Fraction of escaping particles as a function of their initial normalized flux $\psi_\mathrm{norm}$. --- (e) Fraction of particles that escape with energy $E_{kin} \geq 0.05\,$MeV as a function of their initial normalized flux $\psi_\mathrm{norm}$.  }
	\label{fig:Ekin_t}
\end{figure*}


Fig.\ \ref{fig:Ekin_t}(a) shows the kinetic energies of several thousand particles, initialized according to the local Maxwellian distribution, and the mean and the maximum kinetic energy of all 100'000 particles as a function of time. The orbits of a few high energy particles and a few low energy particles have been marked by thick lines to make it easier to follow them.

Starting from the precursor phase, the mean kinetic energy increases (until $t\approx 32.75\,$ms), it then remains constant for a short time-interval (until $t\approx 32.9\,$ms), where-after it slightly drops during the peaking of the ELM, and it finally rises again until the final time. The maximum value of the kinetic energy shows a similar behavior, it just has no phase of decrease during the ELM's peak. Energization is rather moderate, the mean energy increases by a factor of $6$ from  $0.7\,$keV at initial time 
to  $4\,$keV at final time, while the maximum energy increases by a factor of $13$ from the initial $7\,$keV 
to the final $90\,$keV. 
The transient drop in mean energy is clearly associated with an increased loss of particles (with higher-than-average energy) close to the peaking of the ELM, see Fig.\ \ref{fig:Ekin_t}(c), it thus can be interpreted as being caused by the escape of energetic particles. 

The acceleration of the high energy particles is gradual and looks rather systematic (the energy is strictly increasing).  
Some low energy particles' evolution is more reminiscent of a random walk like motion. 

From Fig.\ \ref{fig:Ekin_t}(a) and Fig.\ \ref{fig:Ekin_t}(c), we may conclude that the peak-phase of the ELM is not particularly  associated with particle energization,
but rather with increased particle loss in a short time-window around the peak.

Fig.\ \ref{fig:Ekin_t}(b) and Fig.\ \ref{fig:Ekin_t}(d) indicate that particles are most likely to escape if their initial position is close to the separatrix, largely independent of their initial energy. Fig.\ \ref{fig:Ekin_t}(e) shows that the particles that escape with the highest energies were initialized very close to the separatrix, they are though very few in number. 


\subsection{Acceleration and Heating}

\begin{figure*}[htp]
		\centering
		\includegraphics[width=0.45\linewidth]{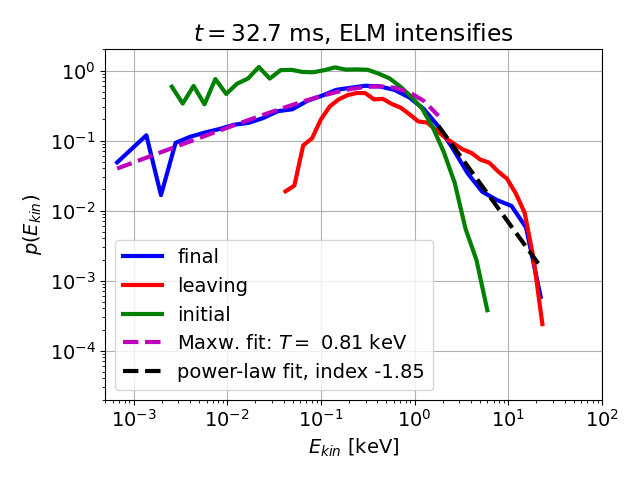}
		\includegraphics[width=0.45\linewidth]{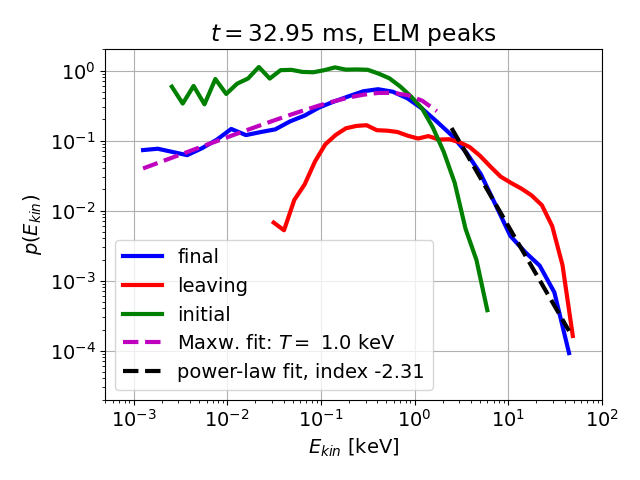}
		\includegraphics[width=0.45\linewidth]{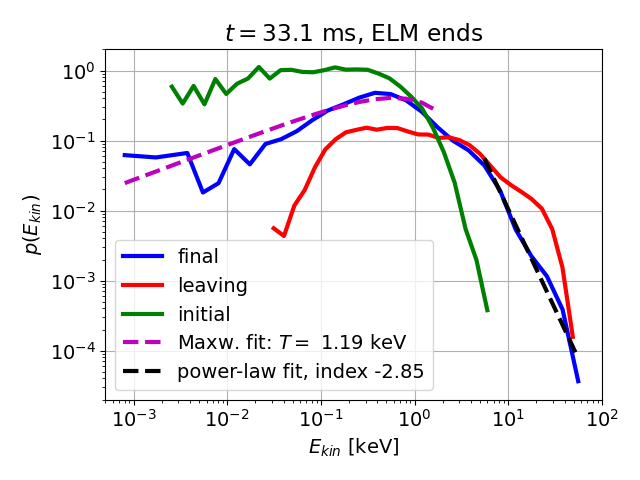}
		\caption{
			Kinetic energy distribution at initial and final time, together with the kinetic energy distribution of the particles that have left up to the time considered, for the three times of interest $t_2,\,t_3,\,t_4$ from Table \ref{tab:toi}. For the distributions of the confined particles, also a Maxwellian fit at low energies and a power-law fit at high energies are shown. }
		\label{fig:p_Ekin}
\end{figure*}


Fig.\ \ref{fig:p_Ekin} shows the kinetic energy distribution at the three times $t_2,\, t_3,\,t_4$ of interest (see Table \ref{tab:toi}), always together with the initial distribution. The figure also shows the distributions of the particles that have left up to the time of interest considered (these distributions are asynchronous, they are based on the particle-energies at the different individual times at which the particles leave). 

Considering first the particles that still are in the system at the times of interest in Fig.\ \ref{fig:p_Ekin}, we find that 
a clear non-Maxwellian tail is formed at the high energies, extending slightly beyond $20\,$keV at $t=t_2$ and reaching almost $50\,$keV at $t=t_3$ and $90\,$keV at $t=t_4$. 
The tail in any case has a part that is of power-law shape, 
most clearly formed at the time of 
the ELM's peak and end,
yet always slightly modulated.
The power-law of the high-energy tail steepens in the course of time, with power-law index -1.9 at $t=t_2\,$, then -2.3 at $t=t_3\,$, and finally -2.9 at $t=t_4\,$.

\begin{figure}[htp]
    \centering
    \includegraphics[width=0.9\linewidth]{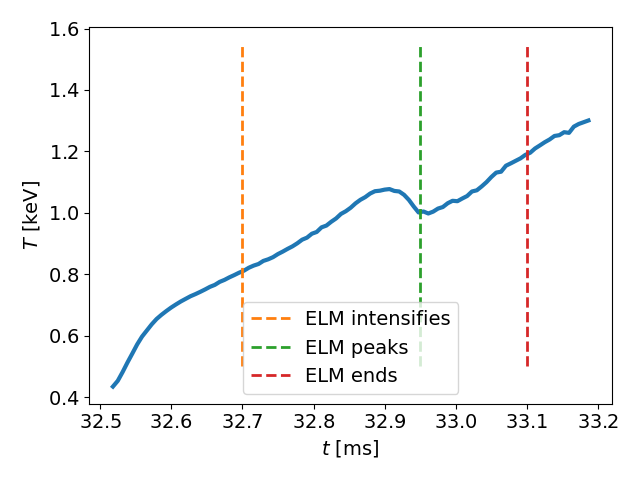}
    \caption{Time evolution of the temperature of the low energy particles (calculated through  Maxwellian fits to the low energy part of the energy distributions). }
    \label{fig:T_t}
\end{figure}

From Fig.\ \ref{fig:p_Ekin} it also follows that
there is gradual heating from initial $0.46\,$keV (see Sect.\ \ref{sec:TP_setup}) to final $1.2\,$keV at $t=t_4$, as a Maxwellian fit at the low energies reveals. Fig.\ \ref{fig:T_t} shows the temperature evolution for the entire duration of the test-particle simulations, and obviously the heating process continues throughout the simulation, with a drop around the peak phase of the ELM, which must be attributed to the increased loss of particles in this phase (as mentioned in Sec.\ \ref{sec:energization}, see Fig.\ \ref{fig:Ekin_t}(c)), covering a wide range of energies, see Fig.\ \ref{fig:lost1} below. We note that  some processes relevant for the temperature evolution have been omitted here, like radiation and collisions (the effect of the latter is discussed in Sec.\ \ref{sect:collisions}).

Turning to the population of escaping particles in Fig.\ \ref{fig:p_Ekin}, we find that (i) there is no clear low energy part of Maxwellian shape of the distributions at any given time, (ii) instead, there is a dominating power-law shape at $t=t_2$, and a double power-law shape at $t=t_3,\,t_4$, with turn-overs at the highest and smallest energies. All power-law parts of the distributions of the leaving particles are flatter than those of the confined particles' distributions.

In sum, we find moderate heating and acceleration of the particles that stay inside, as well as moderate acceleration of the escaping particles. 
 
\subsection{Pitch angle distribution}

\label{sec:pitch-angle}

\begin{figure*}[htp]
    \centering
    \includegraphics[width=0.45\linewidth]{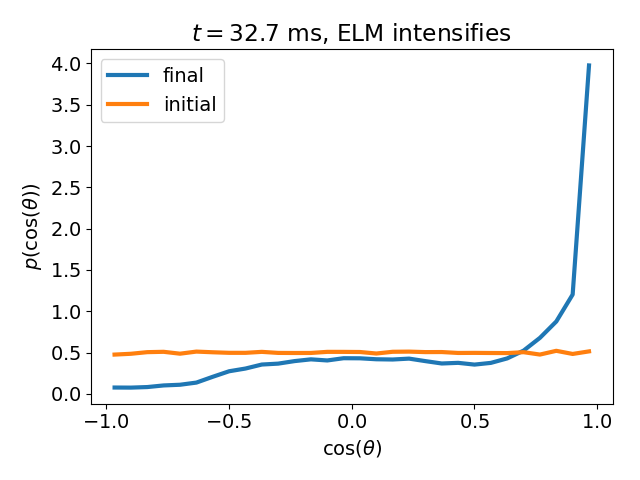}
    \includegraphics[width=0.45\linewidth]{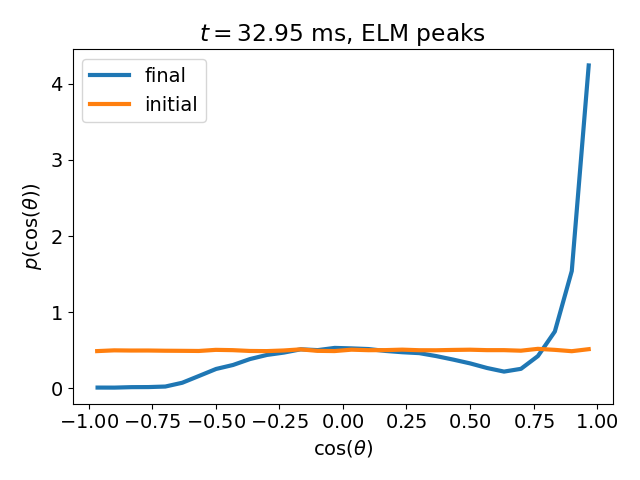}
    \includegraphics[width=0.45\linewidth]{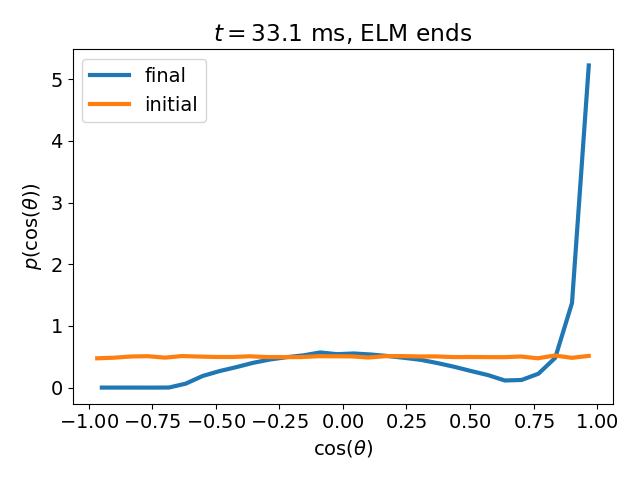}
    \caption{Distribution of $\cos\theta$, with $\theta$ the pitch-angle, for the three different times of interest $t_2, t_3, t_4$, together with the initial distribution. }
    \label{fig:pitch_angle}
\end{figure*}

\begin{figure}[htp]
    \centering
    \includegraphics[width=0.9\linewidth]{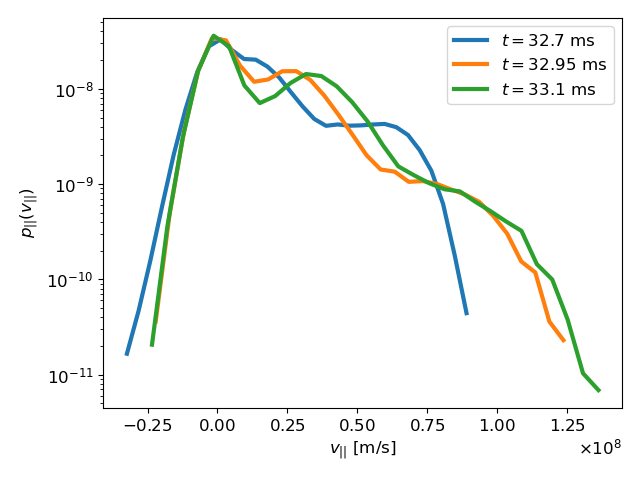}
    \caption{Normalized histogram of the parallel velocity $v_{||}$, for the three times of interest $t_2$, $t_3$, $t_4$ (see Table \ref{tab:toi}).}
    \label{fig:p_vpar}
\end{figure}

The pitch-angle as the angle between the magnetic field at a particle position and the velocity vector is calculated as $\theta = \arctan (u_{\perp}/u_{||})$.
Initially, the direction of the velocity is chosen uniformly random in velocity space, see Sec.\ \ref{sec:TP_setup}, which implies that the initial distribution of $\cos\theta$ is uniform (spherical coordinates are implied).


Fig.\ \ref{fig:pitch_angle} shows the distribution of the cosine of the pitch angle for the three times of interest $t_2,\,t_3,\,t_4$.
Basically at all times, $\cos(\theta) =  1$ or $\theta=0$ is the preferred direction, i.e.\ the particles tend to be aligned with the parallel direction. 
At times $t=t_3,\,t_4$, the perpendicular region around $\theta = \pi/2$ gets slightly repopulated.  
The particles'  preference of the parallel direction is also obvious from the histograms of the parallel velocity $v_{||}$ in Fig.\ \ref{fig:p_vpar}, which show a very clear asymmetry towards positive values. 

As will be confirmed in Sec.\ \ref{the_nature} below, the tendency of the particles to end up with a much higher parallel than perpendicular kinetic energy can be explained by the efficient acceleration in the parallel direction through the parallel electric field, whereby the omission of collisions additionally prevents the possible isotropization of the low-energy particles, see Sec.\ \ref{sect:collisions}. This is a process similar to the acceleration of runaway electrons (REs) during tokamak disruptions, e.g.\ \citet{Breizman2019}. REs are accelerated by the strong toroidal field forming during the current quench and there is evidence that MHD induced field perturbations can also contribute to the formation of a RE seed~\cite{Sommariva2018a}. Disruption induced REs eventually are all moving toroidally in the counter-current direction eventually due to the strong background electric field of the decaying plasma current. In a similar way, we observe electrons here that are primarily accelerated into counter-current direction. 

Magnetic field aligned energy distributions have also been inferred by Ref.~\cite{Freethy2015} from observations of microwave bursts during ELMs at MAST, and it moreover has been shown with PIC simulations that the anomalous Doppler instability causes a rapid isotropization of the distributions that is accompanied by radiation losses due to microwave emission. It is thus plausible that the anomalous Doppler instability would also be triggered and isotropize the distributions in the case of the simulations presented here. However, plasma instabilities are not taken into account in the test-particle code MAGRA.

\subsection{On the nature of the energization process\label{the_nature}}

For a better understanding of the energization process, we now separately analyze and compare the parallel
and the perpendicular energization, and we explore the role of the parallel electric field in the acceleration process.

\begin{figure*}[htp]
    \centering
	\includegraphics[width=0.45\linewidth]{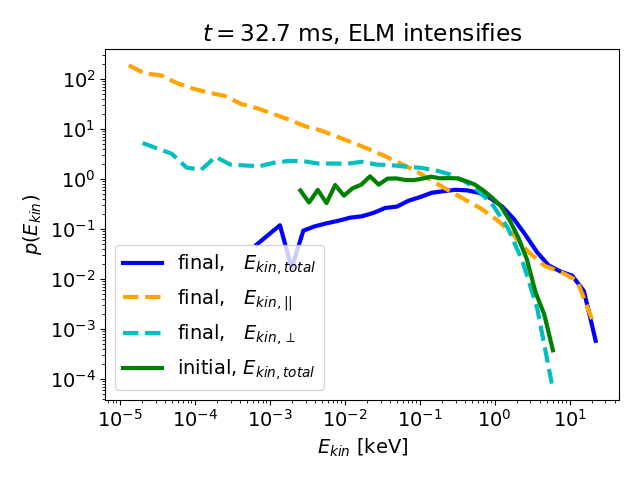}
	\includegraphics[width=0.45\linewidth]{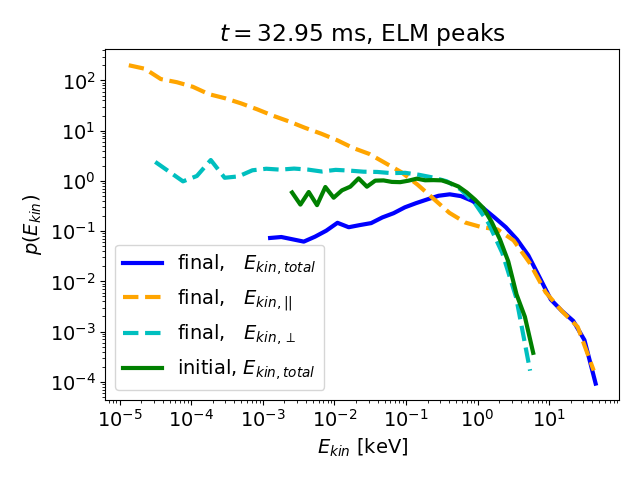}
	\includegraphics[width=0.45\linewidth]{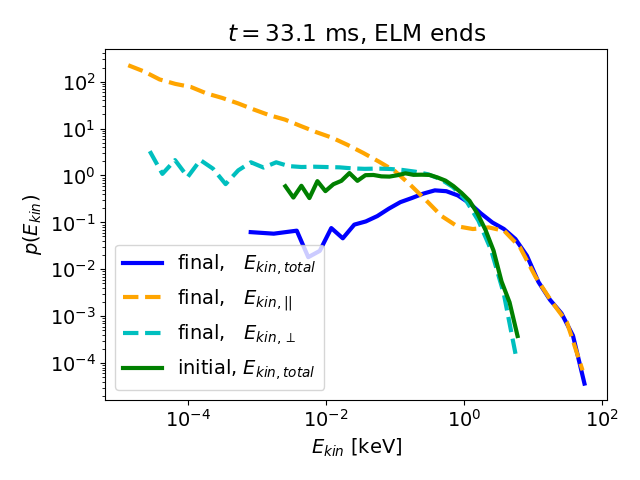}
    \caption{Distribution of the total ($E_{kin,total}$), parallel ($E_{kin,||}$), perpendicular ($E_{kin,\perp}$), and initial ($E_{kin,initial}$) kinetic energy, at the three times of interest $t_2,\,t_3,\,t_4$ (see Table \ref{tab:toi}). }
    \label{fig:Ekin_par_perp}
\end{figure*}

In Fig.\ \ref{fig:Ekin_par_perp}, the  distributions of the total, the parallel, the perpendicular, and the initial kinetic energy are shown at the three times of interest $t_2,\,t_3,\,t_4$.
The distributions of the total and the parallel kinetic energy coincide completely in the tail at the high energies  (small differences must be attributed to the different binning underlying the histograms).
The acceleration process thus is clearly and exclusively in the parallel direction. The form of the guiding center equations (Eqs.\ (\ref{dr_dt}) and (\ref{dupar_dt})) in turn implies that the cause for the acceleration process is the parallel component of the electric field. Magnetic mirror effects only exchange parallel and perpendicular kinetic energy and thus cannot contribute to energizing the particles.

In the low energy range, neither the parallel nor the perpendicular kinetic energy distribution coincide with the total one, which means that heating is a combined effect of the parallel and perpendicular electric field activity. 

\begin{figure*}[htp]
	\centering
	\includegraphics[width=0.45\linewidth]{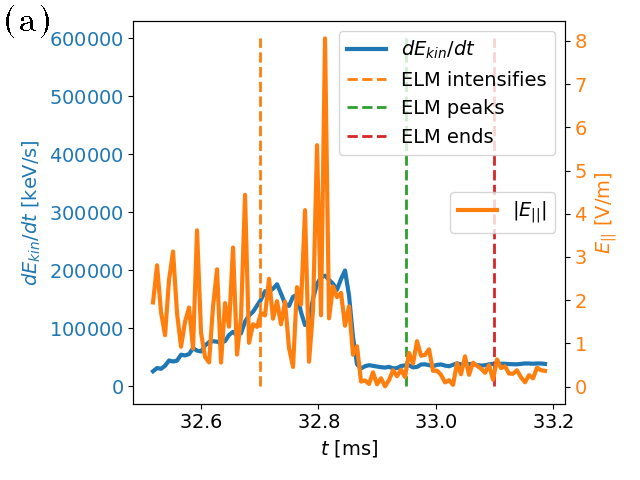}
	\includegraphics[width=0.45\linewidth]{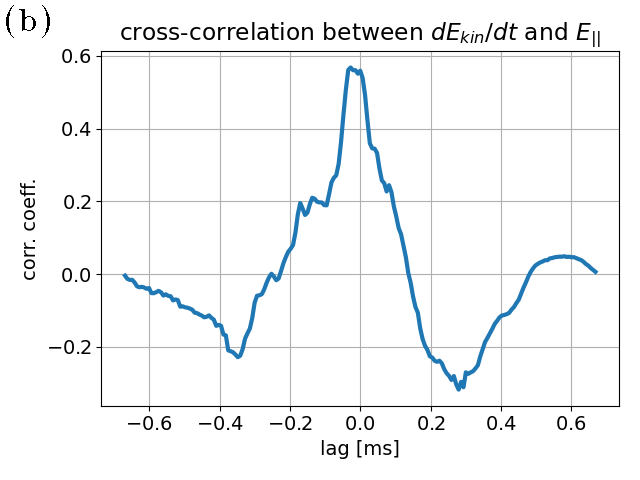}
	 \\
	\includegraphics[width=0.45\linewidth]{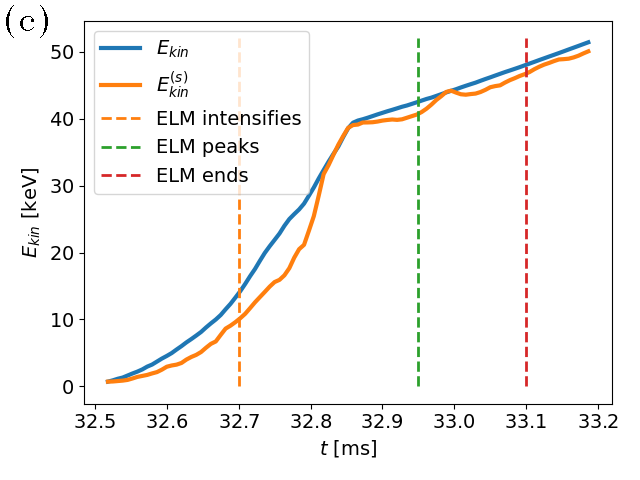}
	\includegraphics[width=0.45\linewidth]{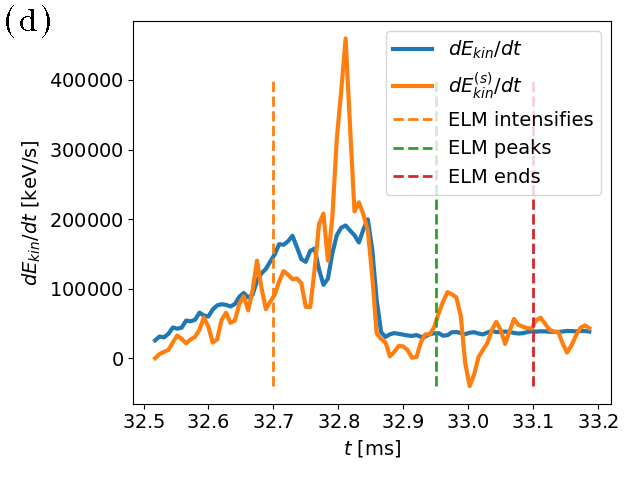}
	\caption{(a) The time derivative of the kinetic energy $dE_{kin}/dt$ and the magnitude of the scaled parallel electric field $\left | E_{||} \right |$ along a particle orbit, for one randomly chosen energetic test-particle. --- (b) Cross-correlation of $dE_{kin}/dt$ and $E_{||}$ in time direction for the same particle. --- (c)
	The kinetic energy $E_{kin}$ along a particle orbit, for one randomly chosen energetic test-particle, in comparison to the kinetic energy $E_{kin}^{(s)}$ yielded by the parallel electric field alone (for details see the text). --- (d) The time derivative of the kinetic energy $dE_{kin}/dt$ along a particle orbit, for one randomly chosen energetic test-particle, in comparison to $dE_{kin}^{(s)}/dt$ as yielded by the parallel electric field alone (for details see the text). 
	}
	\label{fig:corr}
	\end{figure*}

To further explore the question 
whether the acceleration is associated with the passing of particles through regions of enhanced parallel electric fields, we show in Fig.\ \ref{fig:corr}(a) the time derivative $dE_{kin}/dt$ of the kinetic energy as a function of time for one energetic particle, together with the magnitude of the parallel electric field along the orbit. 
Obviously, the systematic increase in energy of the particle is associated with the onset and persistence of large parallel electric fields, yet a more detailed comparison based on this figure is difficult. 

Fig.\ \ref{fig:corr}(b) 
presents the cross-correlation between $dE_{kin}/dt$ and $E_{||}$, for the same  energetic particle. The correlation coefficient of about 0.6 at time-lag almost zero
implies that large values of $dE_{kin}/dt$ are correlated with large values of $E_{||}$, practically without any real time-delay.

A more direct way to investigate the role of the parallel electric field can be achieved by isolating its effect in the equations of motion. We thereto omit all terms from Eq.\ \ref{dupar_dt} except for the parallel electric field, considering the evolution equation
$m_0 du_{||}/dt = q E_{||}^{(s)}(t)$, we  also ignore the spatial evolution according to Eq.\ \ref{dr_dt}, and as $E_{||}^{(s)}(t)$ we use the parallel electric field coarsely sampled along a particle orbit.
We thus numerically integrate  $u_{||}^{(s)}(t)=\int_0^t (q/m_0) E_{||}^{(s)}(t) dt + u_{||,0}$, imposing $u_{||,0}=0$, we then determine the Lorentz factor
$\gamma(t) = \sqrt{1+(u_{||}^{(s)}/c)^2}$
and the kinetic energy
$E_{kin}^{(s)}(t) = (\gamma - 1) m_0 c^2 + E_{kin,0}$, where $E_{kin,0}$ is the actual initial kinetic energy of the test-particle considered. Fig.\ \ref{fig:corr}(c) shows an example of the resulting kinetic energy for one high-energy particle, in comparison to the kinetic energy from the integration of the full equations of motion (as done throughout elsewhere in this article), and Fig.\ \ref{fig:corr}(d) shows the temporal derivatives of these kinetic energies. The alignment between the two time-series is very good, the reduced and simplified integration captures all basic features of the actual energy evolution on a  quantitative level, despite the fact that $E_{||}^{(s)}$ has a substantially lower resolution than the parallel electric field that actually is witnessed by a particle along its orbit. This is a very strong indication that acceleration is exclusively due to the parallel electric field.

We thus have good evidence that the ultimate cause for the acceleration is the parallel electric field, which also is in agreement with the findings in Ref.~\cite{Freethy2015} from observations of microwave bursts and MHD simulations of ELMs at MAST. 

\subsection{Escaping particles}

\begin{figure}[htp]
	\centering
	\includegraphics[width=0.9\linewidth]{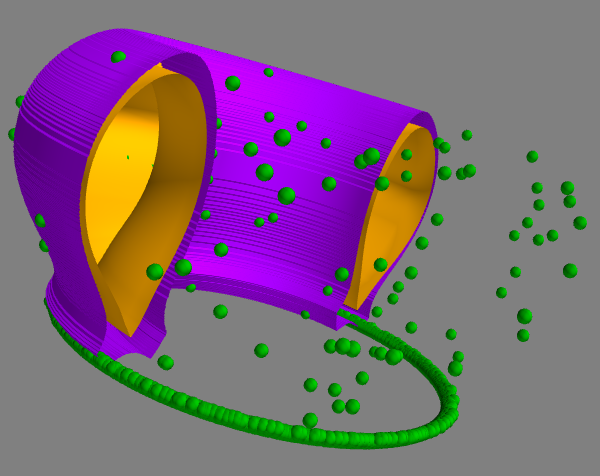}
	\caption{Separatrix (yellow) and plasma boundary (violet), and final position of the escaping particles (green spheres). }
	\label{fig:escape}
\end{figure}

Turning to the escaping particles, we find that 52\% of the particles have left at the final time of interest, $t=t_4$ (the ELM ends), which does basically not change anymore until the end of the simulation (see Fig.\ \ref{fig:Ekin_t}(c)). Fig.\ \ref{fig:escape} shows the last positions of the leaving particles just before they escape (they are stopped still inside the modeling region, see Sec.\ \ref{sec:TP_setup}).
Most of the escaping particles leave in a localized region at the bottom in 
the divertor at the outer leg, 
forming thus one strike line in the divertor. A considerable number of particles also leaves at any height, equally likely at the HFS and the LFS.  

\begin{figure}[htp]
	\centering
	\includegraphics[width=0.9\linewidth]{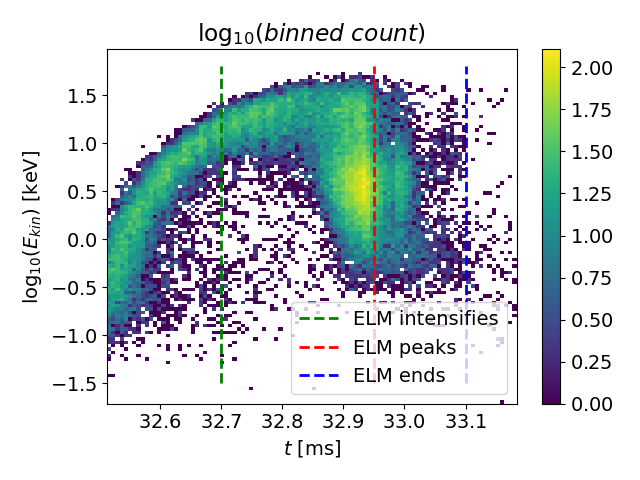}
	\caption{Logarithmic number of escaping particles as a function of logarithmic energy and the time-instant just shortly before they leave the system, calculated through binned statistics along the energy axis.}
	\label{fig:lost1}
\end{figure}

In Fig.\ \ref{fig:lost1}, we show the number of escaping particles as a function of their energy and the  time-instant just shortly before they leave the system. The largest densities of escaping particles appear during the peak phase of the ELM, and they spread over a range of energies from the most energetic particles around 50 keV to basically not energized particles at 0.5 keV (which will have been initialized close to the system boundary), with a peak around 3 keV. There thus is no clearly preferred energy range for the escaping particles during the peak phase of the ELM. In the precursor phase though, the energy of the escaping particles systematically increases in the course of time, from 0.3 keV to 15 keV just before approaching the ELM's peak phase. As in Sec.\ \ref{sec:energization}, we again find that the loss of particles is facilitated in a wide energy range during the ELM's main activity.

\begin{figure}[htp]
	\centering
	\includegraphics[width=0.9\linewidth]{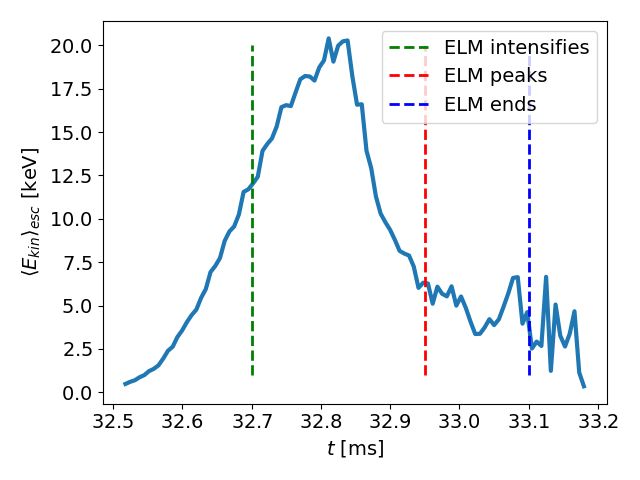}
	\caption{The mean energy of the escaping particles as a function of time.}
	\label{fig:Ekin_mean_escape}
\end{figure}

Fig.\ \ref{fig:Ekin_mean_escape} shows that the mean kinetic energy of the escaping particles is highest in the precursor phase, reaching $20\,$keV, where after it drops to about $5\,$keV at the ELM's peak, and it roughly remains so until the end of the simulation.

\section{Transport in energy space}
\label{sec:transport_energy}

\begin{figure}[htp]
	\centering
	\includegraphics[width=0.9\linewidth]{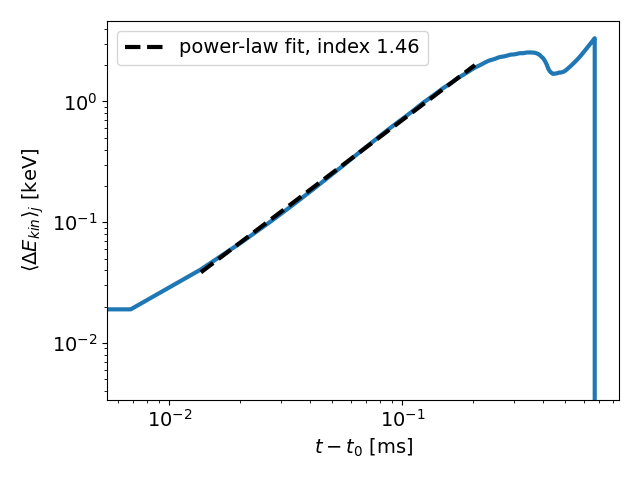}
	\caption{Mean displacement in energy as a function of time.  }
	\label{fig:mdekin}
\end{figure}

\begin{figure}[htp]
	\centering
	\includegraphics[width=0.9\linewidth]{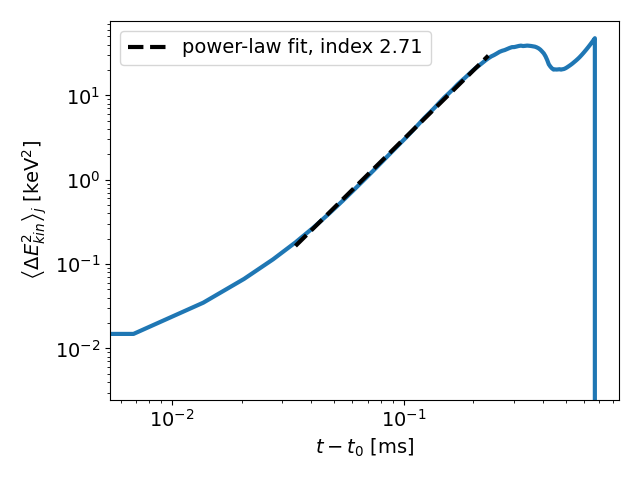}
	\caption{Mean square displacement in energy as a function of time.  }
	\label{fig:msdekin}
\end{figure}

\begin{figure*}[htp]
	\centering
	\includegraphics[width=0.45\linewidth]{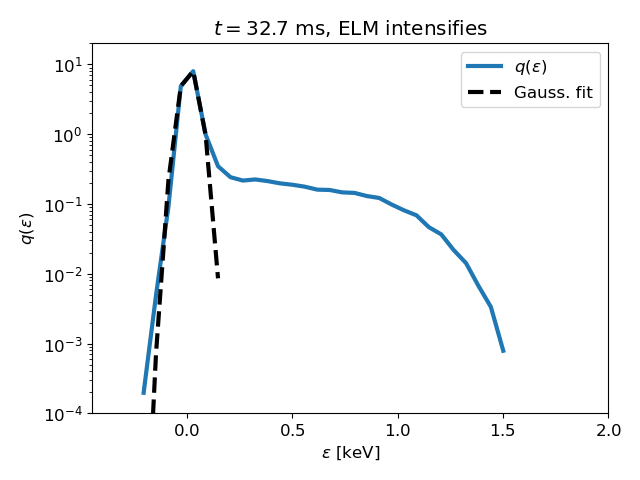}
	\includegraphics[width=0.45\linewidth]{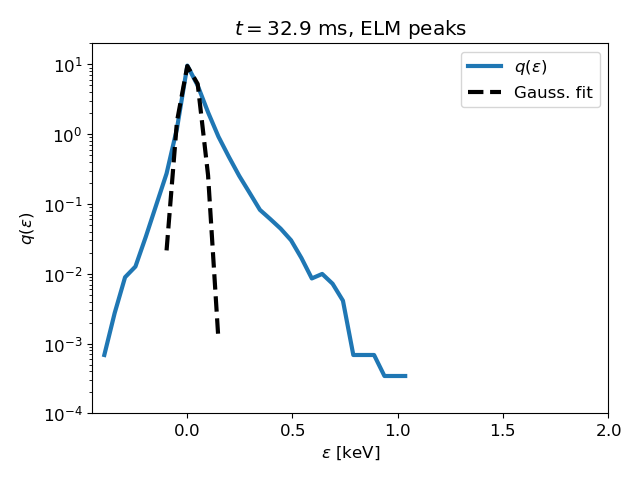}
	\includegraphics[width=0.45\linewidth]{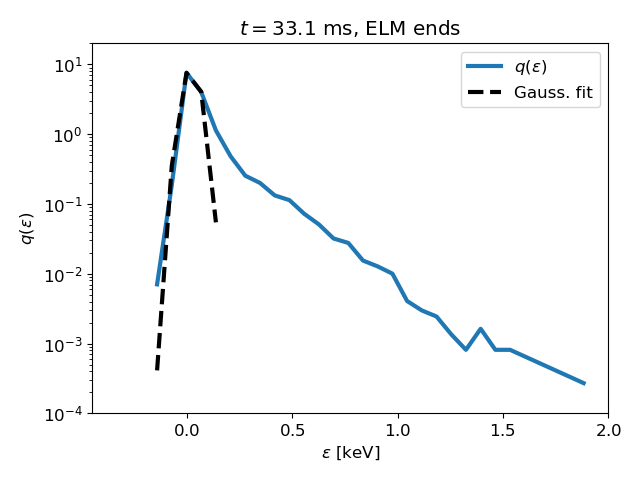}
	\caption{Distribution (normalized histograms) of time-local energy increments $\epsilon_j(t)$ (Eq.\ (\ref{eq:DeltaEkin})), at the three different times $t=t_2,\,t_3,\,t_4$ of interest, together with the fit of a Gaussian distribution.
	  }
	\label{fig:penergyincr}
\end{figure*}

A usual method to characterize transport in energy space is through the mean displacement and the mean square displacement in energy.
The time-global energy increments between the times $t_0$ and $t$ are defined as
\beq
\Delta E_{kin,j}(t) :=  E_{kin,j}(t) - E_{kin,j}(t_0) ,
\label{eq:DeltaEkin_global}
\eeq  
where $j$ is the particle index.

Averaging the global energy increments over the particles yields the mean displacement in energy as a function of time, 
$\left\langle \Delta E_{kin,j}\right\rangle_j(t)$,
which is shown in Fig.\ \ref{fig:mdekin}. In the interval $0.015 < t < 0.2\,$ms, 
a rather clear power-law scaling is formed, with power-law index 1.46.

The mean square displacement in energy is the average of the squared global energy increments, 
$\left\langle \Delta E_{kin,j}^2\right\rangle_j(t)$, and it 
is presented in Fig.\ \ref{fig:msdekin}. Similar to the mean displacement, a power-law scaling 
is found in the time-range $0.03 < t < 0.2\,$ms. The power-law index is $2.71$, implying that transport is super diffusive  and thus clearly of anomalous nature. 
The power-law index of the mean square displacement is about twice as large as the power-law index of the mean displacement, which is a strong hint that transport essentially is of systematic convective nature, and diffusive random walk like motion is of minor importance (see Sec.\ \ref{sec:FP}), which is in accordance with the visual impression from Fig.\ \ref{fig:Ekin_t}(a).

For an estimate of the convective and diffusive transport coefficients in energy, which basically are a kind of derivative of the mean- and mean-square-displacement, 
it is appropriate to consider the time-local energy increments between the times $t-\Delta t$ and $t$, defined as
\beq
\epsilon_{j}(t) :=  E_{kin,j}(t) - E_{kin,j}(t-\Delta t) ,
\label{eq:DeltaEkin}
\eeq  
where $j$ is the particle index. The  time-interval $\Delta t$ should be small enough in order to resolve the involved time-scales of energization, but still large enough so that all particles show a notable change in energy. Here, we use $\Delta t = 7\times 10^{-3}\,$ms, a fraction of 100 of the total simulation time. This approach follows~\cite{Ragwitz2001}, and it takes into account that the transport coefficients could be time-dependent.

Valuable information on the energization process is given by the
distribution of the time-local energy increments $\epsilon_{j}(t)$, Eq.\ (\ref{eq:DeltaEkin}), 
which are shown in
Fig.\ \ref{fig:penergyincr} 
for the three times of interest.
All distributions are of Gaussian shape at low energies, with clear exponential tails at positive large energies,
and during the peak phase of the ELM there also is an exponential tail at negative energies, which is of smaller extent than the tail at positive energies. 
The asymmetric bias towards positive values of the distributions of increments 
is in accordance with the fact that acceleration prevails.  
After all, the energization process clearly obeys non-Gaussian statistics.

\subsection{Remarks on the Fokker-Planck equation in energy space}\label{sec:FP}

\begin{figure*}[htp]
	\centering
	\includegraphics[width=0.45\linewidth]{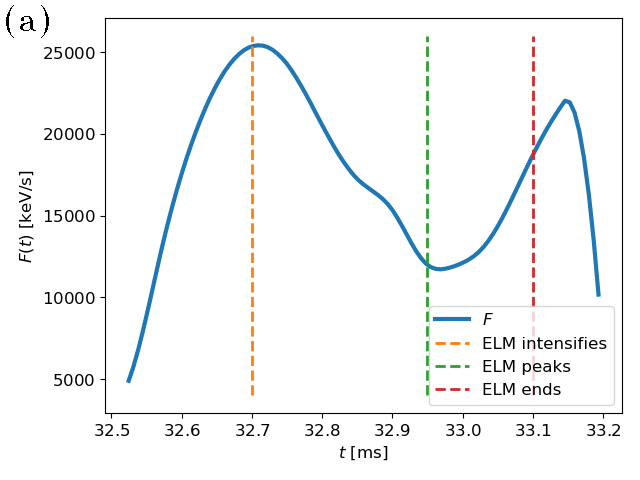}
	\includegraphics[width=0.45\linewidth]{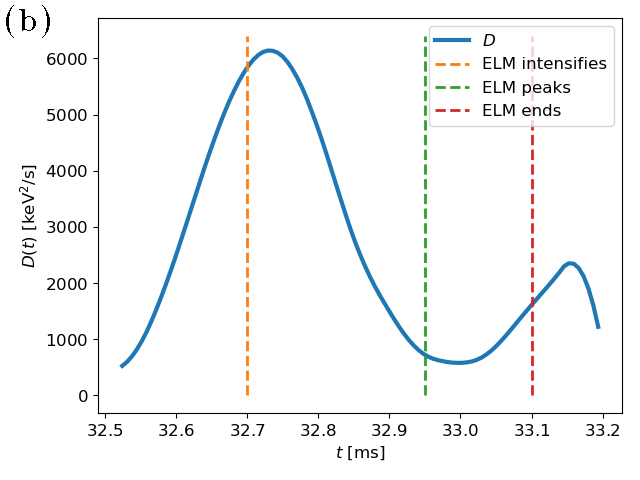}
	\\
	\includegraphics[width=0.45\linewidth]{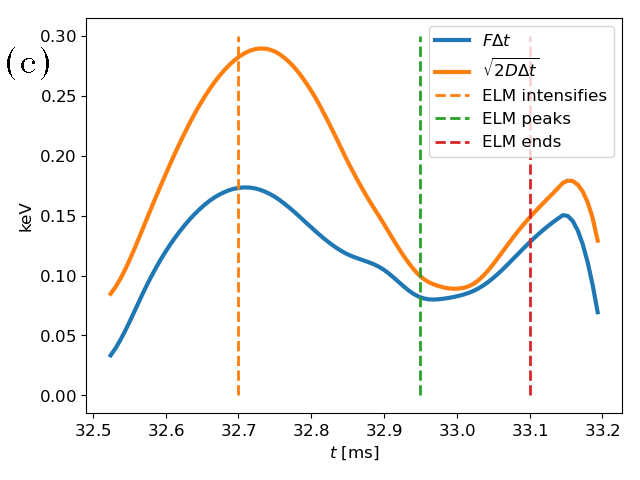}
	\caption{(a) Convection coefficient $F$ and (b) diffusion coefficient $D$ as a function of time for the high energy particles with $E_{kin} \geq 2\,$keV, and (c) comparison of the convective and the diffusive term in Eq.\ (\ref{eq:stoch}).  }
	\label{fig:F_and_D}
\end{figure*}

The convection and the diffusion coefficient, $F$ and $D$, respectively, are defined as
\beq
F=\left\langle \epsilon_{j}\right\rangle_j(t)/\Delta t ,
\label{eq:F}
\eeq
\beq
D=\left\langle \epsilon_{j}^2\right\rangle_j(t)/(2\Delta t) , \label{eq:D}
\eeq
with $\Delta t$ constant and adopting the same value as in the definition of the time-local energy differences $\epsilon_j$, Eq.\ (\ref{eq:DeltaEkin}),
see e.g.~\cite{Ragwitz2001}.

For finite time-intervals $\Delta t$, the value of $D$ can be contaminated e.g.\ by the square of the drift coefficient $F$ (see e.g.~\cite{Ragwitz2001}), so that, instead of using Eq.\ (\ref{eq:D}), we follow 
\citet{Ragwitz2001}
and 
we calculate $D$ from the expression
\beq
\left\langle\epsilon_{j}^2\right\rangle_j(t)
= 2\Delta t \, D + \Delta t^2  F^2   ,
\label{eq:D_corrected}
\eeq 
Here, 
$F$ is assumed to be determined correctly through Eq.\ (\ref{eq:F}).

The estimates of $F$ and $ D $ as a function of time are shown in Fig.\ \ref{fig:F_and_D}(a) and (b) for the high energy particles with $E_{kin}\geq 2\,$keV. In the course of time, there are some modulations in the estimates by a factor of 5 in the case of $F$, and by a factor of 12 in the case $D$, whereby the local maxima and minima of $ F $ and $ D $ are in-phase. In particular, $F$ and $D$ have a local minimum at the ELM's peak.

Eqs.\ (\ref{eq:F}) and (\ref{eq:D_corrected}) represent empirical estimates of the transport coefficients, as they enter the Fokker-Planck equation of the form
\beq
\frac{\p p}{\p t}  = \frac{\p}{\p E_{kin}}\left[\frac{\p \left(Dp\right)}{\p E_{kin}}  -Fp \right] -p/\tau_{esc},
\eeq
with $p$ the kinetic energy distribution, and $-p/\tau_{esc}$ the loss term.
The Fokker-Planck equation (without the loss term) in turn corresponds to the stochastic differential equation
\beq
dE_{kin,t} = F dt + \sqrt{2Ddt} N_t ,
\label{eq:stoch}
\eeq
where $N_t$ is Gaussian white noise with mean zero
and variance 1, and $dt$ is a small time-step,
see e.g.~\cite{Gardiner2009}. Eq.\ (\ref{eq:stoch}) makes clear that $F$ represents systematic, deterministic energization, while $D$ gives rise to stochastic and normally diffusive displacements in energy.  
 
The relative importance of the two transport coefficients is revealed when, based on Eq.\ (\ref{eq:stoch}), comparing the values of $F dt$ to $\sqrt{2Ddt}$, for, say, $dt\approx \Delta t$, which is shown in Fig.\ \ref{fig:F_and_D}(c).
The diffusive term is about 2 times larger than the convective term when the ELM intensifies, and the two terms roughly  equalize on from about the ELM's peak until its end, which suggests that systematic, convective energization is of equal importance as stochastic transport
for the energetic particles
in the main phase of the ELM. We though note that 
\citet{Friedrich2002} propose even more finite time interval corrections in the estimation of $D$ 
than we applied in Eq.\ (\ref{eq:D_corrected}), which are likely to further change the shape and relative importance of $D$.

The estimates of the transport coefficients $F$ and $D$ and the use of the Fokker-Planck equation are though slightly inconsistent for the following reasons:
The fact that the energy increments follow a distribution with exponential tails, allowing relatively large increments, is in conflict with the smallness assumption for the energy increments that is made in the theoretical derivation of the Fokker-Planck equation, see e.g.~\cite{Gardiner2009}. The distributions of increments are in principle also incompatible with the stochastic differential equation (Eq.\ (\ref{eq:stoch})) that is equivalent to the Fokker-Planck equation, and where the explicitly stated stochastic term is Gaussian white noise. 
Yet, as stated, the theoretical inconsistency is of a slight nature, and it is not obvious whether the Fokker-Planck approach still can be successful or will break down in the case considered here, since it is not a clear case of non-local, fractional transport, as e.g.\ in \cite{Isliker2017a}). We thus consider it worthwhile to explore the applicability of the Fokker-Planck equation by implementing the transport coefficients that we determined here numerically, we though leave this for future work.






\section{The effect of collisions}
\label{sect:collisions}

\begin{figure}[htp]
    \centering
    \includegraphics[width=0.9\linewidth]{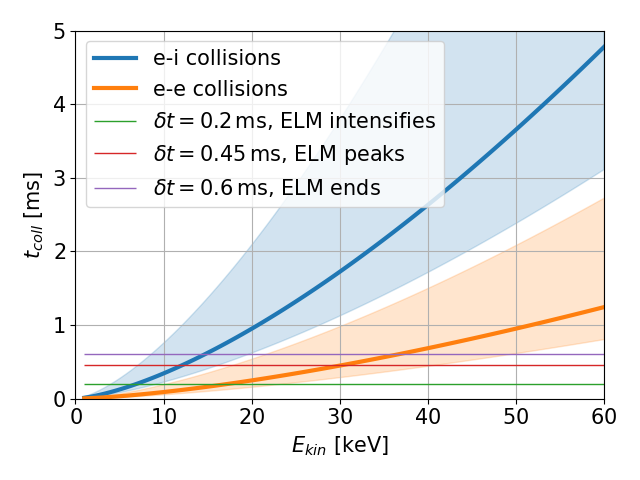}
    \caption{Collision times as a function of particle energy, for electron-ion (e-i) and electron-electron (e-e) collisions. The shaded regions indicate the variation of the collision times with the particle density from its maximum ($7\times 10^{19}\,$m$^{-3}$) to its minimum ($2\times 10^{19}\,$m$^{-3}$) value in the pedestal, while the solid lines show the collision times for the mean pedestal density ($4.5\times 10^{19}\,$m$^{-3}$). }
    \label{fig:coll_time}
\end{figure}

Collisions have not be taken into account in the test-particle simulations, and here we estimate their potential effect on the presented results. The electrons can collide with background particles that are either thermal ions (deuterium, as it is assumed in the MHD simulations) or thermal electrons, both with a temperature of $0.46\,$keV (see Sect.\ \ref{sec:TP_setup}) and a density  varying from $2\times 10^{19}\,$m$^{-3}$ to $7\times 10^{19}\,$m$^{-3}$
in the region from which the spatial initial conditions are chosen, see Sect.\ \ref{sec:TP_setup}.

According to Sect.\ \ref{sec:energization} and Fig.\ \ref{fig:Ekin_t}, the main energization takes place between the initial time $t_0$ and the time $t_3$ of the ELM's peak.
At the three times of interest $t_2,\, t_3,\, t_4$, (see Table \ref{tab:toi}), collisional effects will have acted over a time interval $\delta t$ measured on from $t_0=32.5\,$ms, which yields $\delta t = 0.2$, $0.45$, $0.6\,$ms, respectively, so the main energization phase lasts about $0.45\,$ms.

The collision times $t_{coll}$ as a function of particle energy are shown in Fig.\ \ref{fig:coll_time} for electron-electron and electron-ion collisions (based on the momentum loss or slowing down Coulomb collision frequency for a particle of a given speed in the Lorentz collision model, see e.g.\ \citet{Huba2018}).
The distribution of the low-energy particles forms a heated Maxwellian at all times and up to about $1\,$keV  (see Fig.\ \ref{fig:p_Ekin}). In this energy range, the collision times are smaller than the energization time-interval $\delta t$ for all times of interest, as Fig.\ \ref{fig:coll_time} shows, so a non-negligible number of collisions would take place, and the heating at low energies would be affected, namely reduced by the attempt of the collisions to thermalize the electrons back to the initial $0.46\,$keV.  

At energies above $20\,$keV, the electron-ion collision times, as well as the electron-electron collision times for particles located beyond the separatrix, are larger than the energization time-intervals $\delta t$ for all times of interest (see Fig.\ \ref{fig:coll_time}), implying that the high energy end of the tail of the energy distributions would not be affected by collisions, which in particular holds true for the escaping particles. 
Particles confined within the separatrix will be unaffected by electron-ion collisions at energies above $20\,$keV, and by electron-electron collisions at energies above $40\,$keV, which concerns the very high end of the energy distributions. 

In sum, the highest energies reached by the test-particles that we report here would not be affected by collisions. Also, the high energy end of the tail of the energy distributions for the particles outside the separatrix, including the escaping particles, would maintain its shape, while for particles within the separatrix also the tail's shape must be expected to be altered by collisions to some degree, not though its extent. 
We plan to explicitly include collisions and to study their effects in future kinetic investigations of particle energization during ELMs.

\section{Conclusions}
\label{sec:conclusions}

We have investigated the heating and acceleration of electrons during an eruptive ELM event on the kinetic level, whereby we also shed light on the nature of the energization process, and we analyzed transport in energy space in detail. Moreover, we performed a statistical analysis of the MHD data per se, in order to better understand the physical mechanisms for the phenomena of heating and acceleration that we find.

The filaments, traced through the parallel electric field, consist of a set of helically winded tube-like structures in the vicinity of the separatrix, which are aligned with the magnetic field and appear at both, the HFS and the LFS, being most intense during the peak-phase of the ELM. 
The filaments' topological structure is very simple, they are line-like at small scales and clearly do not form a fractal.
    
The probability distributions of the parallel electric field exhibit power-law tails and thus they are clearly non-Gaussian in shape, which definitely plays a crucial role in the phenomena of particle acceleration that we observe during the filament eruption.

The energization of the test-particles takes place in a short time-window (with duration $0.5\,$ms) in the phase where the ELM intensifies. The high-energy particles are moderately accelerated, forming a non-thermal tail, partly of power-law shape, and reaching energies up to $90\,$keV. The low-energy particles are gradually heated from $0.4$ to $1.2\,$keV. As we have demonstrated, for the low energy thermal particles collisions would play a role and lead to less heating than what we find here in the absence of collisions. The high energy end of the non-Maxwellian tail, on the other hand, would not be affected by collisions.

The majority of the particles escape during the energization phase, at an increased rate close to the peak of the ELM. Their kinetic energy distributions do not show Maxwellian parts at low energies and rather are of single or double power-law shapes. The escapes preferably take place at one line-like location in the outer divertor leg.  The appearance of several rather discrete strike-lines in the divertor region has also been observed in other MHD simulations of ELMs for JET~\cite{Huijsmans2017}, as well as in experimental measurements of the ELM energy fluence profile in the divertor~\cite{Eich2017}. The high energy particles have the signature of  runaway electrons, given that they almost all escape to the wall, and they form a non-Maxwellian tail, whose high energy end is not affected by collisions.  
We note that also increased losses of fast ions have been reported from particle tracing in the electromagnetic fields of ELM simulations done with JOREK, which moreover were confirmed in experimental measurements at several tokamaks~\cite{Garcia-Munoz2013a},~\cite{Garcia-Munoz2013b}.

Acceleration takes place exclusively in the parallel direction, and it actually solely is the parallel electric field that causes the acceleration (this finding is in agreement with observations of microwave bursts and MHD simulations of ELMs in Ref.~\cite{Freethy2015}). Heating, on the other hand, is a mixture of parallel and perpendicular energization, in particular all perpendicular drifts, such as e.g.\ the $E$-cross-$B$ drift, only contribute to heating. As a consequence, the kinetic energy distributions are anisotropic during the intense energization, the particle velocities are predominantly field aligned, clearly preferring the parallel over the anti-parallel direction. The anisotropy in principle might give rise to plasma instabilities (which are not included in our simulations), likely the anomalous Doppler instability causing  radiation losses due to microwave emission and  isotropization of the distributions on fast time-scales \heinz{($\sim 0.05\,\mu$s)} \cite{Freethy2015}. 

The acceleration of the high energy particles is systematic and a gradual process, indicating that particles and large $E_{||}$ regions remain in phase for a long time (at least for certain particles). The mean-square displacement in energy indicates clearly super-diffusive behavior, and the transport coefficients 
show that systematic  transport of convective nature 
is of equal importance as
stochastic random walk like diffusive  features (for the energetic particles in the main phase of the ELM).

Some specific characteristics of heating and acceleration met here have occurred also in astrophysical applications, e.g.~\cite{Isliker2017b}, and most notably in the case of magnetic flux emerging from the solar  convection zone into the solar corona~\cite{Isliker2019}, so to say the astrophysical analogue of filamentary eruptions from the core into the edge region in magnetically confined plasmas. Common features are: 
(1) there is heating at low energies; 
(2) a non-thermal tail of power-law shape is formed in the kinetic energy distributions, albeit here at moderate energies;  
(3) acceleration 
is caused exclusively by the parallel electric field; 
(4) collisions lead to reduced heating but cannot affect the high-energy end of the non-Maxwellian tail in its shape.

Effects not accounted for in this study include particle collisions,  which are discussed in Sect.\ \ref{sect:collisions}, "re-filling" of particles in the $\psi_{norm}=0.8\dots1$ range by particles that are initially further inwards or outwards, creation of new electrons by ionization of deuterium atoms or impurities, and certain two-fluid and/or kinetic effects that could affect the evolution of the MHD fields. Nevertheless, the study presented here gives a detailed look at the processes related to electron acceleration during ELMs. Future studies can take into account additional effects to further refine the picture regarding the accelleration, resulting kinetic instabilities, and on the long-term self-consistent evolution of plasma and non-thermal particles. 

Issues not addressed in this study but planned to be tackled in future work include: (i) The moderate acceleration that the electrons undergo is a combined convective and diffusive process, implying that a classical Fokker-Planck type approach 
should be appropriate for the modeling of transport in energy space. It would be worthwhile to verify that this indeed is the case. (ii) It certainly would be of interest to also quantify and characterize radial transport. 
(iii) As shown in Sec.\ \ref{sect:collisions}, collisional effects may be of importance at the lower energies, and we plan to explicitly include them in future kinetic studies.

\textbf{Acknowledgements:} 

This work has been carried out within the framework of the EUROfusion Consortium, funded by the European Union via the Euratom Research and Training Programme (Grant Agreement No 101052200 - EUROfusion). Views and opinions expressed are however those of the author(s) only and do not necessarily reflect those of the European Union or the European Commission. Neither the European Union nor the European Commission can be held responsible for them. Some of the work was done with the TSVV project number 8 on MHD transients. Some of the simulations were performed by the Marconi-Fusion supercomputer operated by CINECA.


\bibliographystyle{unsrtnat}
\bibliography{filaments,vlahosturb2}

\end{document}